\documentclass[review, 12pt]{elsarticle}
\usepackage{lineno,hyperref}
\usepackage{setspace}
\usepackage[justification=centering]{caption}
\usepackage{subfigure}
\usepackage{amssymb}
\usepackage{color}
\usepackage{amsmath}
\usepackage{nomencl}
\usepackage{array}
\usepackage{hyperref}
\usepackage{multirow}
\usepackage{geometry}
\usepackage{algorithm}
\usepackage[noend]{algpseudocode}
\usepackage{makeidx}
\usepackage{float}
\DeclareMathOperator{\E}{\mathbb{E}}

\DeclareMathOperator{\Var}{Var}

\makeindex
\providecommand{\makenomenclature}{\makeglossary}
\usepackage{etoolbox}
\makenomenclature
\def\bm{\boldsymbol}
\journal{Journal of \LaTeX\ Templates}

\begin{document}

\begin{frontmatter}

\title{Momentum-based Accelerated Mirror Descent Stochastic Approximation for Robust Topology Optimization under Stochastic Loads}

\author[address1]{Weichen Li}
\author[address1,address2]{Xiaojia Shelly Zhang \corref{mycorrespondingauthor}}
\cortext[mycorrespondingauthor]{Corresponding author}
\ead{zhangxs@illinois.edu}

\address[address1]{Department of Civil and Environmental Engineering, University of Illinois Urbana-Champaign, 205 North Mathews Ave, Urbana, IL 61801, USA}
\address[address2]{Department of Mechanical Science and Engineering, University of Illinois Urbana-Champaign}

\begin{abstract}
Robust topology optimization (RTO) improves the robustness of designs with respect to random sources in real-world structures, yet an accurate sensitivity analysis requires the solution of many systems of equations at each optimization step, leading to a high computational cost. To open up the full potential of RTO under a variety of random sources, this paper presents a momentum-based accelerated mirror descent stochastic approximation (AC-MDSA) approach to efficiently solve RTO problems involving various types of load uncertainties. The proposed framework can perform high-quality design updates with highly noisy stochastic gradients. We reduce the sample size to two (minimum for unbiased variance estimation) and show only two samples are sufficient for evaluating stochastic gradients to obtain robust designs, thus drastically reducing the computational cost. We derive the AC-MDSA update formula based on $\ell_1$-norm with entropy function, which is tailored to the geometry of the feasible domain. To accelerate and stabilize the algorithm, we integrate a momentum-based acceleration scheme, which also alleviates the step size sensitivity. Several 2D and 3D examples with various sizes are presented to demonstrate the effectiveness and efficiency of the proposed AC-MDSA framework to handle RTO involving various types of loading uncertainties.
\end{abstract}

\end{frontmatter}

%\linenumbers

\section{Introduction}
Topology optimization has been widely used in many disciplines, such as aerospace engineering \cite{Aage2017,Zhu2016}, biomedical engineering \cite{Sutradhar2010,Challis2010}, and architectural design \cite{Beghini2014}. The main goal of topology optimization is to find the distribution of material to achieve optimized performance \cite{Bendsoe2004,Sigmund2013}. While the classical setting of topology optimization assumes problem-related parameters that are deterministic, real-world structures are subjected to various sources of randomness, such as load, material property, and geometry, which can influence the layout of optimized designs. Thus, robust topology optimization (RTO) has been employed to improve the robustness of designs concerning random sources \cite{Sigmund2009,Wang2011,Asadpoure2011,Dunning2013,Kogiso2008,Zhao2014,Chen2010,Zhao2015,Chan2019,Keshavarzzadeh2017,Wu2016,Cardoso2019}.

% RTO with load randomness: importance, existing approaches and their limits, and gap statement(Cited papers are mostly on RTO with load randomness)
One common random source comes from the loading, which includes load magnitudes, directions, locations, and distributions. Many studies have contributed to the RTO with load randomness using various approaches, such as semidefinite programming \cite{Ben-Tal1997}, conversion to many load cases \cite{Guest2008,Dunning2011}, first-order reliability method approximation \cite{Kogiso2008}, Karhunen-Loeve expansion to model stochastic load fields \cite{Zhao2014,Chen2010},  perturbation techniques \cite{daSilva2018}, stochastic collocation \cite{Zhao2015}, univariate dimension reduction \cite{Chan2019}, polynomial chaos expansion \cite{Keshavarzzadeh2017}, game theory\cite{Holmberg2017}, linear elastic theory \cite{Zhao2014_2}, and non-probabilistic interval uncertainty \cite{Wu2016}.
These approaches successfully produce robust optimized designs. For large-scale problems, particularly in three dimensions (3D), some may require a relatively high computational cost as they typically solve multiple systems of equations at each optimization step in order to accurately estimate the sensitivity information. In this work, we aim to reduce the computational cost associated with RTO problems using stochastic approximation. %

%However, compared to the deterministic setting, these RTO approaches still demand multiple times more computational cost due to solving multiple linear systems or eigenvalue problems. Such levels of computational cost are still challenging when solving large-scale three-dimensional problems, which resembles real-world structures. Therefore, reducing the computational cost and improving scalability is crucial to RTO.

% Introduction of MDSA
Stochastic approximation (SA) \cite{Robbins1951} is a family of stochastic optimization methods known for its low computational cost and effectiveness \cite{Nemirovski1983}. In the standard setting, SA methods solve stochastic optimization problems with the objective function in the form of the expectation of a stochastic function \cite{Nemirovski2009}. Instead of computing the exact gradient, the classic SA method uses a stochastic one as the gradient descent direction. Thus, SA is also known as stochastic gradient descent (SGD). The SGD was initially developed by Robbins and Monro \cite{Robbins1951} and improved in \cite{Nemirovski2009,Nemirovski1983,Polyak1992}. In \cite{Nemirovski1983,Nemirovski2009}, the classic SA (or SGD) is generalized to the mirror descent stochastic approximation (MDSA) by replacing the traditional $\ell_2$-norm definition of distance in SGD with a more general definition. With the general setting, MDSA adapts its update to the underlying geometry of the feasible space and obtain improvements in the convergence performance \cite{Nemirovski2009,Lan2012b,Zhang2020}. One of the most popular versions of MDSA is the entropic MDSA, which is based on the $\ell_1$-norm setting \cite{Nemirovski2009}. 
%
%While the one-sample SGD may seem counterintuitive, it is proved mathematically that in the case of a convex cost function, the SA algorithm converges to the optimum \cite{Robbins1951,Nemirovski2009}.
%
In a related area, smooth convex programming, accelerated methods (also known as momentum methods) were first developed by Polyak \cite{Polyak1964} and significantly improved by Nesterov \cite{Nesterov1983,Nesterov2018}. These methods are referred to as the accelerated gradient descent and proved to possess an unimprovable rate of convergence for convex problems as a linear Krylov subspace method \cite{Nesterov2018}. The accelerated methods are incorporated into SA and MDSA to speed up the convergence of stochastic optimization\cite{Lan2012,Ghadimi2012,Ghadimi2013,Ghadimi2016}. Inspired by a popular version of accelerated SA methods, accelerated mirror descent stochastic approximation (AC-MDSA) \cite{Lan2012}, this paper derives an AC-MDSA framework tailored for the topology optimization accounting for stochastic loads. 

% Brief background of MDSA
%In the standard setting, the SA method solves a stochastic optimization problem with the objective being the expectation of a stochastic cost function \cite{Nemirovski2009}. The classic SA algorithm mimics the procedure of the projected gradient descent method \cite{Nemirovski2009}, but it uses the stochastic gradient computed based on only one sample as opposed to the many samples in Monte Carlo-based methods. While the one-sample algorithm may seem counterintuitive, it is proved mathematically that in the case of a convex cost function, the stochastic procedure converges to the optimum \cite{Robbins1951,Nemirovski2009}.

In the field of topology optimization, the idea of integrating stochastic optimization algorithms has been recently explored in a few studies. For instance, Zhang et al.\cite{Zhang2017} proposed a stochastic sampling algorithm that requires $5$ to $6$ samples to estimate the gradient and solve deterministic topology optimization problems with hundreds of load cases. De et al. \cite{De2019} applied SGD algorithms to compliance minimization of RTO problems with load uncertainty and shows improvements over GCMMA \cite{Svanberg2007}. Pflug et al. \cite{Pflug2020} developed a continuous stochastic gradient method (CSG) that shows superiority over traditional SGD methods when applied to the expected compliance minimization (without the variance term). Both \cite{De2019} and \cite{Pflug2020} treat the volume constraint as a penalization term in the objective function, thereby converting the constrained optimization to an unconstrained problem. The volume constraint represents a feasible domain bounded a plane in which $\ell_1$-norm-based entropic MDSA performs better than the $\ell_2$-norm-based SGD (and its variants) \cite{Nemirovski2009}. Recently, the $\ell_1$-norm-based entropic MDSA has been proposed and tailored for topology optimization with many deterministic load cases \cite{Zhang2020} and requires only a single sample at each optimization step, thereby significantly reducing the computational cost compared to the standard weighted average formulation \cite{Zhang2020}. Theoretical and numerical comparisons of the entropic MDSA and SGD are also carried out therein, and show better performance (objective function values and computational time) of the $\ell_1$-norm entropic MDSA than SGD ($\ell_2$-norm-based) for compliance minimization with a volume constraint \cite{Zhang2020}. The advantage of entropic MDSA comes from the use of $\ell_1$-norm and entropic distance function to mimic the underlying geometry of the feasible design space represented by the linear volume constraint \cite{Zhang2020}. Therefore, we focus on the entropic MDSA with the $\ell_1$-norm setting in this study. 

%Specifically, we use the Accelerated Mirror Descent Stochastic Approximation (AC-MDSA) \cite{Lan2012}, which is one of the most recent versions of SA algorithms. The method is tailored to the RTO problem with load randomness. As illustrated in later sections, the AC-MDSA can reduce the computational cost of solving RTO problems to the level of solving deterministic problems by sampling only twice at each optimization step.
%
%
%This study uses the AC-MDSA method, one of the most recent versions of accelerated SA methods, to solve RTO problems, which can be treated as stochastic optimization problems. The method requires only two samples to compute the stochastic gradient at each optimization step, reducing the computational cost to the level of solving deterministic problems. Through numerical examples, we show that the two-sample AC-MDSA method can produce designs with various levels of robustness with statistical consistency and possesses extremely high efficiency. According to the best of our knowledge, this is the first work that integrates the AC-MDSA algorithm with RTO problems.

In this work, we propose a novel momentum-based AC-MDSA algorithm to solve RTO problems with the volume constraint involving various types of loading uncertainties. The proposed AC-MDSA approach can perform high-quality design variable updates with noisy stochastic gradients. As a result, we demonstrate that only two samples are sufficient for computing the stochastic gradients at each optimization step, which is the minimum number of samples for unbiased variance estimation. %When used together with an iterative solver for the state equation, the proposed algorithm allows for a relaxed solver tolerance without sacrificing the accuracy of optimized results, leading to additional computational efficiency.
Second, in order to adapt to the underlying geometry of the feasible set defined by the volume constraint, we derive the explicit update formula in the $\ell_1$-norm setting by introducing the entropy function as the distance-generating function in the AC-MDSA method. Third, we present adaptive step-size recalibration and damping schemes which, in conjunction with the momentum-based acceleration mechanism, to improve the convergence performance of AC-MDSA with significantly reduced sensitivity to various step size choices. Through numerical examples in both 2D and 3D, we showcase that the proposed AC-MDSA approach can efficiently produce robust designs with respect to different types of loading uncertainties and exhibits scalable performance for RTO problems of various problem sizes and geometries.

%The classic SA algorithm relies on the Euclidean description of distance, and the MDSA generalizes the classic SA by using a general description of distance \cite{Nemirovski2009}. Such generalization allows for choosing "proper" distance functions to fit the problem at hand, which in this study, is the RTO problem.

% Organizatin of this paper
The remainder of this paper is organized as follows. Section 2 reviews the RTO formulation for compliance minimization problem considering various load uncertainties. Section 3 introduces the theoretical background of AC-MDSA and derives a momentum-based entropic AC-MDSA update algorithm for the RTO problem. Section 4 proposes algorithmic techniques for improving convergence performance, including adaptive step size recalibration and damping schemes. Section 5 presents four numerical examples illustrating the effectiveness and efficiency of the proposed entropic AC-MDSA algorithm in producing robust optimized designs under various loading uncertainties. Finally, Section 6 provides concluding remarks.

%%%%%%%%%%%%%%%%%%%%%%%%%%%%%%%%%%%%%%%%%%%%%%% END of Section 1 %%%%%%%%%%%%%%%%%%%%%%%%%%%%%%%%%%%%%%%%%%%%%%%%%%%%%%%%%%%%%%%%%%%%%%%%

\section{Robust topology optimization formulation}

In this section, the RTO formulation of compliance minimization problem considering loading uncertainties is introduced, and the unbiased estimations of the objective function and gradient of the RTO formulation are presented using a finite number of samples. In this work, we focus on the density-based approach \cite{Bendsoe2004}.

For a given mesh consisting of $n$ finite elements, the RTO aims to minimize the weighted sum of the mean and variance of the compliance under load randomness \footnote{A similar approach used by many studies is the weighted sum of mean and standard deviation as the objective function, this work focuses on the weighted sum of mean and variance. %because obtaining an unbiased estimation of variance is relatively straightforward.
}. More specifically, the RTO formulation is introduced as follows:
\begin{equation}\label{eq:RTO original form.}
\begin{split}
& \min_{\boldsymbol{x}} J(\boldsymbol{x}) =
\frac{\kappa}{w} \E\left[C(\boldsymbol{x},\boldsymbol{\xi})\right] + \frac{1-\kappa}{w^2} \Var \left[C(\boldsymbol{x},\boldsymbol{\xi})\right] \\
& \text{s.t.} \quad \frac{ V\left(\boldsymbol{x} \right)}{V_0} - V_f = 0 \\
& \quad \quad \, x^{(i)} \in [0,1], \quad i=1,2,...,n \\
& \text{with} \, \boldsymbol{K}\left(\boldsymbol{E}\left(\boldsymbol{x}\right)\right) \boldsymbol{u}(\bm{x},\bm{\xi}) = \boldsymbol{f}(\boldsymbol{\xi}),
\end{split}
\end{equation}

\nomenclature{$J$}{RTO objective function}
\nomenclature{$\kappa$}{Relative weight of mean and variance}
\nomenclature{$w$}{Normalization factor of the RTO's objective function}
%\nomenclature{$\mu [\cdot]$}{Mean}
\nomenclature{$\Var [\cdot]$}{Variance}
%\nomenclature{$\sigma [\cdot]$}{Standard deviation}
\nomenclature{$\boldsymbol{\xi}$}{Random vector}
%$$ = \alpha \E_{\boldsymbol{\xi}}[C(\boldsymbol{x},\boldsymbol{\xi})] + \beta ( \E_{\boldsymbol{\xi}} [C^2(\boldsymbol{x},\boldsymbol{\xi})] - \E_{\boldsymbol{\xi}}^2 [C(\boldsymbol{x},\boldsymbol{\xi})] )	$$
%$$ s.t. \quad \boldsymbol{l}^T \boldsymbol{x} - v \leq 0 $$
%$$ x^{(i)} \in [0,1], \quad i=1,2,...,n $$
%
\noindent
where $\bm{x}$ is the design variable vector; $\textbf{\textit{f}}(\bm{\xi})$ is the random load vector with $\bm{\xi}$ being a random vector representing various types of load uncertainty; $\bm{K}$ and $\bm{u}$ are the global stiffness matrix and displacement vector, respectively; $V_0$ is the total volume of the design domain; and $V_f$ is the prescribed volume fraction. For a given structure with design variable $\bm{x}$, $V(\bm{x})$ stands for the total volume of that structure as follows,
\begin{equation}
V\left(\boldsymbol{x}\right) = \sum_{i=1}^{n} v^{(i)} \bar{x}^{(i)} = \boldsymbol{v}^T \bar{\boldsymbol{x}} = \boldsymbol{v}^T \boldsymbol{H} \boldsymbol{x},
\end{equation}
where  $v^{(i)}$ and $\bar{x}^{(i)}$ is the volume and the filtered/physical density of the $i$th element, respectively; and $\boldsymbol{H}$ is the matrix representation of density filter \cite{Bourdin2011,Zhu2016}, such that $\bar{\bm{x}}=\bm{H}\bm{x}$, which is used to prevent the checkerboard pattern and achieve mesh-independent designs \cite{Diaz1995,Jog1996,Sigmund1998}. In addition, the modified simplified isotropic material with penalization (SIMP) \cite{Bendsoe2004,Bendse1989,Sigmund2007} is adopted, which interpolates the Young's modulus of each element as 
\begin{equation} \label{eq:modified SIMP}
E^{(i)}\left(\bar{x}^{(i)} \left(\boldsymbol{x}\right)\right) = E_{min} + \left(\bar{x}^{(i)}\left(\boldsymbol{x}\right) \right)^p \left(E_0 - E_{min}\right), \quad i = 1,...,n,
\end{equation}
where $E_0$ is the Young's modulus of the solid material; $E_{min}$ is the Ersatz stiffness which is taken to be $10^{-4}$; and $p$ is the SIMP penalization parameters, which is taken to be $3$ \cite{Andreassen2010} in this study. 
The objective function of the RTO formulation \eqref{eq:RTO original form.} is a weighted sum of the expectation and the variance of the compliance, $C(\boldsymbol{x},\boldsymbol{\xi}) = \boldsymbol{f} ^T (\boldsymbol{\xi}) \boldsymbol{u}(\boldsymbol{x},\boldsymbol{\xi})$, where $ \E\left[\cdot \right]$ and $\Var\left[\cdot \right]$ stand for expectation and variance operators, respectively; and $\kappa\in[0,1]$ is a prescribed coefficient representing the relative importance of the expectation over the variance in the objective function. Because the expectation and the variance of the compliance have different units, we normalize their relative weights by $w$ and $w^2$, respectively, where  $w = \bar{\boldsymbol{f}} ^T\bar{\boldsymbol{f}} / E_0$ with $\bar{\boldsymbol{f}} = \E [\boldsymbol{f}(\boldsymbol{\xi})]$ being the expectation of the load vector $\bm{f}(\bm{\xi})$ \cite{Dunning2013}.

%\subsection{Density-based deterministic topology optimization: compliance minimization}
%
%This work employs the density-based approach with the modified Simplified Isotropic Material with Penalization (SIMP) model \cite{Sigmund2007}. The density filter \cite{Zhu2016,Bourdin2011} is used to prevent the checkerboard pattern and achieve mesh-independent designs \cite{Diaz1995,Jog1996,Sigmund1998}. Based on this setting, the nested formulation of the compliance minimization problem is: 
%%
%\begin{equation}\label{eq:TO form.}
%\begin{split}
%& \min_{\boldsymbol{x}} C\left(\boldsymbol{x}\right) = \boldsymbol{f}^T \boldsymbol{u}\left(\boldsymbol{x}\right)\\
%& \text{s.t.} \quad \frac{ Vol\left(\boldsymbol{x} \right)}{V_0} - V_f \leq 0 \\
%& \quad \quad \, x^{(i)} \in [0,1], \quad i=1,2,...,n \\
%& \text{with} \, \boldsymbol{K}\left(\boldsymbol{E}\left(\boldsymbol{x}\right)\right) \boldsymbol{u} = \boldsymbol{f}
%\end{split}
%\end{equation}

%
\nomenclature{$\boldsymbol{x}$}{Design variable vector}
\nomenclature{$\bm{x}^*$}{Optimized solution} 
\nomenclature{$C$}{Compliance}
\nomenclature{$\boldsymbol{f}$}{Load vector}
\nomenclature{$\boldsymbol{u}$}{Displacement vector}
\nomenclature{$Vol$}{Total material volume}
\nomenclature{$V_0$}{Volume of design domain}
\nomenclature{$V_f$}{Volume fraction}
\nomenclature{$\boldsymbol{K}$}{Global stiffness matrix}
\nomenclature{$\boldsymbol{E}$}{Vector of element Young's modulus}
%
%\noindent
%where $\boldsymbol{x}$ is the design variable vector, $C$ is the compliance, $\boldsymbol{f}$ is the load vector, $\boldsymbol{u}$ is the displacement vector, $V$ is total volume of material, $V_0$ is the domain volume, $V_f$ is the prescribed volume fraction, $\boldsymbol{K}$ is the global stiffness matrix, and $\boldsymbol{E}$ is the vector of the Young's moduli of all elements. The total volume of material $V\left(\boldsymbol{x}\right)$ has the following expression:
%$$
%V\left(\boldsymbol{x}\right) = \sum_{i=1}^{n} v^{(i)} \bar{x}^{(i)} = \boldsymbol{v}^T \bar{\boldsymbol{x}} = \boldsymbol{v}^T \boldsymbol{H} \boldsymbol{x},
%$$

\nomenclature{$v^{(i)}$}{Element volume of the $i$th element}
\nomenclature{$\boldsymbol{v}$}{Element volume vector}
\nomenclature{$\bar{x}^{(i)}$}{Physical density of the $i$th element}
\nomenclature{$\bar{\boldsymbol{x}}$}{Physical density vector}
\nomenclature{$\boldsymbol{H}$}{Density filter matrix}
%
%\noindent
%where $v^{(i)}$ is the volume of element $i$, $\bar{x}^{(i)}$ is the filtered/physical density, and $\boldsymbol{H}$ is the matrix representation of density filter. Based on the modified SIMP model, the Young's modulus is:
%\begin{equation} \label{eq:modified SIMP}
%E^{(i)}\left(\bar{x}^{(i)} \left(\boldsymbol{x}\right)\right) = E_{min} + \left(\bar{x}^{(i)}\left(\boldsymbol{x}\right) \right)^p \left(E_0 - E_{min}\right), \quad i = 1,...,n,
%\end{equation}

\nomenclature{$E^{(i)}$}{Young's modulus of the $i$th element}
\nomenclature{$E_{min}$}{Young's modulus of void}
\nomenclature{$E_{0}$}{Young's modulus of solid}
\nomenclature{$p$}{Penalization parameter}
The stochastic gradient of the objective function $J$ in formulation \eqref{eq:RTO original form.} is given by 
\begin{equation}
\bm{g}(\bm{x})\doteq\nabla_{\bm{x}}J(\bm{x})=\E[\bm{G}(\bm{x},\bm{\xi})]=\E\Big[\frac{\kappa}{w}\bm{G}^{\mu}(\bm{x},\bm{\xi})+\frac{1-\kappa}{w^2}\bm{G}^{\Var}(\bm{x},\bm{\xi})\Big],
\end{equation}
where
\begin{equation}
\bm{G}^{\mu}(\bm{x},\bm{\xi})\doteq\nabla_{\bm{x}}C(\bm{x},\bm{\xi})\quad \text{and}\quad\bm{G}^{Var}(\bm{x},\bm{\xi})\doteq 2\Big(C(\bm{x},\bm{\xi})-\E[C(\bm{x},\bm{\xi})]\Big)\nabla_{\bm{x}}C(\bm{x},\bm{\xi}),
\end{equation}
respectively. 
%\begin{equation}
%\begin{aligned}
%&\bm{g}_{\mu}(\bm{x})\doteq\nabla_{\bm{x}}\Big(\E[C(\bm{x},\bm{\xi})]\Big)=\E[\nabla_{\bm{x}}C(\bm{x},\bm{\xi})]\quad \text{and}\\
%&\bm{g}_{Var}(\bm{x})\doteq\nabla_{\bm{x}}\Big(\Var[C(\bm{x},\bm{\xi})]\Big)=\E[C(\bm{x},\bm{\xi})\nabla_{\bm{x}}C(\bm{x},\bm{\xi})]-\E[C(\bm{x},\bm{\xi})]\E[\nabla_{\bm{x}}C(\bm{x},\bm{\xi})],
%\end{aligned}
%\end{equation}
%respectively. 
%
In the above expressions, the stochastic gradient of the compliance with respect to the design variable, $\nabla_{\bm{x}}C(\bm{x},\bm{\xi})$, is obtained through the chain rule as 
\begin{equation}
\nabla_{\bm{x}}C(\bm{x},\bm{\xi})=\bm{H}^T\nabla_{\tilde{\bm{x}}}C(\bm{x},\bm{\xi}),
\end{equation} 
where $\nabla_{\bar{\bm{x}}}C(\bm{x},\bm{\xi})$ is the stochastic gradient of the compliance with respect to the filtered design variable $\bar{\bm{x}}$, whose $i$th component is given by 
\begin{equation}
\frac{\partial C(\bm{x},\bm{\xi})}{\partial \bar{x}^{(i)}}=-p(\bar{x}^{(i)})^{p-1}\Big(\bm{u}^{(i)}(\bm{x},\bm{\xi})\Big)^T\bm{k}_0^{(i)}\bm{u}^{(i)}(\bm{x},\bm{\xi})
\end{equation}
Symbols $\boldsymbol{u}^{(i)}$ and $\boldsymbol{k_0}^{(i)} $ are the nodal displacement vector and the element stiffness matrix (corresponds to solid material) of the $i$th element, respectively. 

%%---- then introduce unbiased estimates 
In this work, we employ unbiased estimators of the objective function and its stochastic gradient. The unbiased estimators of $\E[C(\boldsymbol{x},\boldsymbol{\xi})]$ and $\Var [C(\boldsymbol{x},\boldsymbol{\xi})]$ using $m$ samples are denoted by ${\mu}_{m}(\bm{x})$ and ${\Var}_{m}(\bm{x})$ with ${\Var}_{m}(\bm{x})=({\sigma}_{m} (\bm{x}))^2$, where ${\sigma}_{m} (\bm{x})$ is the estimate of the standard deviation. The estimators ${\mu}_{m}(\bm{x})$ and ${\Var}_{m}(\bm{x})$ are given by 
\begin{equation}
\begin{aligned}
&{\mu}_{m}(\bm{x}) = \frac{1}{m} \sum_{j=1}^{m} C(\boldsymbol{x},\boldsymbol{\xi}_j) \quad \text{and}\\
&{\Var}_{m}(\bm{x}) = \frac{1}{m-1} \sum_{j=1}^{m} \Big(C(\boldsymbol{x},\boldsymbol{\xi}_j) - {\mu}_{m}(\bm{x})\Big)^2,
\end{aligned}
\end{equation}
respectively, where $\boldsymbol{\xi}_j, j=1,...,m$ are independent and identically distributed (i.i.d.) samples of the random vector $\boldsymbol{\xi}$. Notice that, for the variance estimator ${\Var}_{m}(\bm{x})$, it requires $m \geq 2$. Accordingly, the unbiased estimator of objective function $J(\bm{x})$, denoted by ${J}_m(\bm{x})$, is given by
\begin{equation}\label{estimated_obj}
{J}_{m}(\boldsymbol{x}) =\frac{\kappa}{w} {\mu}_{m}(\bm{x}) + \frac{1-\kappa}{w^2}{\Var}_{m}(\bm{x})
\end{equation}
\nomenclature{${J}_m$}{Sample-based RTO objective function}
\nomenclature{${\mu}_{m} [\cdot]$}{Unbiased estimate of mean using $m$ samples}
\nomenclature{${\Var}_{m} [\cdot]$}{Unbiased estimate of variance using $m$ samples}
\nomenclature{${\sigma}_{m} [\cdot]$}{Estimate of standard deviation using $m$ samples}
\nomenclature{$\hat{J}$}{Estimate of RTO objective function value using $m=10,000$ samples}
\nomenclature{$\hat{\mu} [\cdot]$}{Unbiased estimate of mean of compliance using $m=10,000$ samples}
\nomenclature{$\hat{\sigma} [\cdot]$}{Estimate of standard deviation of compliance using $m=10,000$ samples}
\nomenclature{$m$}{Number of samples}
\nomenclature{$c_1$}{Weighting factor of estimated mean}
\nomenclature{$c_2$}{Weighting factor of estimated variance}
Similarly, the unbiased estimators of $\nabla_{\bm{x}}\Big(\E[C(\bm{x},\bm{\xi})]\Big)$ and $\nabla_{\bm{x}}\Big(\Var[C(\bm{x},\bm{\xi})]\Big)$ using $m$ samples, which are respectively denoted by ${\bm{G}}_m^{\mu}(\bm{x})$ and ${\bm{G}}_m^{\Var}(\bm{x})$, take the forms of 
\begin{equation}
\begin{aligned}
&{\bm{G}}_m^{\mu}(\bm{x})=\frac{1}{m}\sum_{j=1}^{m}\nabla_{\bm{x}}C(\bm{x},\bm{\xi}_j)\quad\text{and}\\
&{\bm{G}}_m^{\Var}(\bm{x})=\frac{2}{m-1}\Bigg\{\sum_{j=1}^{m}\Big( C(\boldsymbol{x},\boldsymbol{\xi}_j) \nabla_{\boldsymbol{x}} C(\boldsymbol{x},\boldsymbol{\xi}_j) \Big) - {\mu}^{C}_{m}(\bm{x}){\bm{G}}_{m}^{\mu}(\bm{x})  \Bigg\}
\end{aligned}
\end{equation} 
Accordingly, the corresponding unbiased estimator of the stochastic gradient of the objective function $\bm{g}(\bm{x})$ using $m$ samples, which is later denoted as ${\bm{G}}_{m}$ takes the form of
\begin{equation}\label{eq: sensitivity}
{\bm{G}}_{m}(\bm{x})=\frac{\kappa}{w}{\bm{G}}_m^{\mu}(\bm{x})+\frac{1-\kappa}{w^2}{\bm{G}}_m^{\Var}(\bm{x})
\end{equation}
We note that, as required by ${\bm{G}}_{m}^{\Var}(\bm{x})$, at least $two$ i.i.d. samples are needed to evaluate the above unbiased gradient estimator, namely $m\geq 2$. 

%\textbf{Suggestion: something which may increase the mathematical flavor of the paper, is to add an appendix, which show that the estimators: $\mu^{C}_{m}(\bm{x})$, $\text{Var}^{C}_{m}(\bm{x})$, ${\textbf{G}}^{\mu}_{m}(\bm{x})$ and ${\textbf{G}}^{\text{Var}}_{m}(\bm{x})$ are in fact unbiased. For example, you can follow the similar procedures in this reference \href{https://en.wikipedia.org/wiki/Bias_of_an_estimator}{here}. }

%Denote $c_1 = \kappa / w$ and $c_2 = \left(1-\kappa\right)/w^2$, the $m$-sample estimator of the objective is:
%%
%\begin{equation}\label{eq:RTO original form m-sample.}
%\min_{\boldsymbol{x}} {J}_m(\boldsymbol{x}) =
%c_1 {{\mu}}_{\boldsymbol{\xi}}\left[C(\boldsymbol{x},\boldsymbol{\xi})\right] + c_2 {{\Var}_{m}}_{\boldsymbol{\xi}} \left[C(\boldsymbol{x},\boldsymbol{\xi})\right], \\
%\end{equation}
\nomenclature{$\boldsymbol{G}_m$}{Stochastic gradient with respect to design variables estimated with $m$ samples}
\nomenclature{$\nabla$}{Gradient operator}

We remark that, if we use the unbiased gradient estimator \eqref{eq: sensitivity} together with the commonly used design update schemes in topology optimization, % such as the Optimality Criteria (OC) method and the Method of Moving Asymptotes (MMA) \cite{Svanberg1987},
a large sample size $m$ is needed. This is because those update algorithms typically require higher accuracy in the estimation of gradient \eqref{eq: sensitivity} to perform high-quality updates \cite{Zhang2020}, which leads to a large sample size $m$ and the solution of $m$ linear systems (in the limit of $m\rightarrow\infty$, we have ${\bm{G}}_{m}(\bm{x})\rightarrow \bm{g}(\bm{x})$). Thus, the associated computational cost can be prohibitive, particularly for large-scale problems. 

To address this challenge, we propose an accelerated MDSA algorithm in Section 3 tailored for the RTO formulation \eqref{eq:RTO original form.}. Compared with the standard optimization algorithms in topology optimization, AC-MDSA is a stochastic optimization method, which can perform high-quality design variable update with highly noisy gradient estimations. As we demonstrate in the design examples, with the tailored AC-MDSA method proposed in this work, we can efficiently and accurately solve RTO problems with only $two$ samples (i.e., $m=2$) at every optimization step, where $2$ is the minimum sample size for the unbiased gradient estimator.

%
%requiring only a small sample size while still leading to convergence. This requirement can be met by stochastic approximation (SA) methods \cite{Robbins1951}, which, in its standard setting, uses only one sample and is proved to be convergent for convex stochastic optimization problems \cite{Nemirovski2009}.
%
%
%In the next section, we introduce an advanced version of SA methods, called Accelerated Mirror Descent Stochastic Approximation (AC-MDSA) \cite{Lan2012}. This method, with some modifications, will serve as the engine to solve the RTO problem \eqref{eq:RTO original form.} efficiently.

%%%%%%%%%%%%%%%%%%%%%%%%%%%%%%%%%%%%%%%%%%%%%%% END of Section 2 %%%%%%%%%%%%%%%%%%%%%%%%%%%%%%%%%%%%%%%%%%%%%%%%%%%%%%%

\section{Accelerated Mirror Descent Stochastic Approximation: theory and algorithm}
This section introduces the background of AC-MDSA and derives the update algorithm when applied to the RTO problem. We first review the general framework of the MDSA \cite{Nemirovski1983,Nemirovski2009} and introduce an accelerated MDSA using momentum-based techniques. One major advantage of the MDSA is that, through its general definition of the distance-generating function, the design variable update can be adapted according to the underlying geometry of the feasible set (see \cite{Nemirovski2009,Lan2012b} for detailed discussions). Exploiting this advantage, this section then derives the update formula of the AC-MDSA in the $\ell_1$-norm setting with the entropy function as the distance-generating function.

\subsection{Mirror descent stochastic approximation (MDSA)}
The MDSA is introduced in \cite{Nemirovski2009} to solve stochastic optimization problems of the form
\begin{equation}\label{eq:SA_Obj}
\min_{\boldsymbol{x} \in X} \big\{ \phi\left(\boldsymbol{x}\right)=  \E \left[{\Phi}\left(\boldsymbol{x},\boldsymbol{\xi}\right)\right]\big\},
\end{equation}

\nomenclature{$\phi\left(\cdot\right)$}{General objective function}
\nomenclature{$X$}{Nonempty bounded convex set}
\nomenclature{$\E$}{Expectation operation}
\nomenclature{$\Phi\left(\cdot,\cdot\right)$}{General stochastic function}
\noindent
where $X \subset \mathbb{R}^n$ is the feasible set of $\bm{x}$ (typically assumed to be a nonempty bounded convex set), and $\bm{\xi}$ is a random vector with a given probability distribution. 
%It is further assumed that the expectation and differentiation of \eqref{eq:SA_Obj} is interchangeable. This holds when ${\Phi}(\cdot,\boldsymbol{\xi})$ is convex, and $\phi{\left(\cdot\right)}$ is finite valued in a neighborhood of $x$ \cite{Strassen1965,Nemirovski2009}. 
The gradient of the objective function is given by:
\begin{equation}\label{eq:grad_s}
\nabla\phi\left(\boldsymbol{x}\right)=\E \left[\nabla_{\boldsymbol{x}} {\Phi}\left(\boldsymbol{x},\boldsymbol{\xi}\right)\right]=\E \left[\boldsymbol{G}\left(\boldsymbol{x},\boldsymbol{\xi}\right)\right],
\end{equation}
\nomenclature{$\boldsymbol{G}\left(\cdot,\cdot\right)$}{Stochastic gradient}
\noindent
where $\boldsymbol{G}\left(\boldsymbol{x},\boldsymbol{\xi}\right)=\nabla_{\boldsymbol{x}} {\Phi}\left(\boldsymbol{x},\boldsymbol{\xi}\right)$ is the stochastic gradient. We note that, although we assume the differentiability of ${\Phi}\left(\boldsymbol{x},\boldsymbol{\xi}\right)$ with respect to $\boldsymbol{x}$, the above setting is applicable to the non-smooth case \cite{Nemirovski2009}.

% NOT FINISHED YET!!!!!!!!!!!!!!!
%MDSA can be seen as a proximal stochastic gradient method, in which the prox-function can be generalized to any strongly convex functions associated with a strongly convex distance-generating function. .... [Not Finished yet!!!!]
% NOT FINISHED YET!!!!!!!!!!!!!!!

Before we introduce the general framework of MDSA, let us first introduce the relevant notations \cite{Nemirovski2009}. We denote $\|\cdot\|$ as a generalized norm defined on $\mathbb{R}^n$ with $\|\boldsymbol{x}\|_*\doteq \sup_{\|\boldsymbol{y}\|\leq1} \boldsymbol{y}^T\boldsymbol{x}$ being its dual norm. We define $\omega(\cdot): X \rightarrow \mathbb{R}$ as a distance-generating function with modulus $\alpha > 0$ with respect to norm $\|\cdot\|$, such that $\omega(\cdot)$ is convex and continuous on $X$, continuously differentiable, and strongly convex with parameter $\alpha$ with respect to  $\|\cdot\|$, namely,
\begin{equation}\label{eq:dist_gen}
\left(\boldsymbol{x}'-\boldsymbol{x}\right)^T \left(\nabla\omega\left(\boldsymbol{x}'\right)-\nabla\omega
\left(\boldsymbol{x}\right)\right)\geq\alpha \|\boldsymbol{x}'-\boldsymbol{x}\|^2 \quad \forall \ \boldsymbol{x}', \boldsymbol{x} \in X
\end{equation}
\nomenclature{$\omega\left(\cdot\right)$}{Distance generating function}
\nomenclature{$\boldsymbol{x}'$}{General variable vector}
\nomenclature{$\boldsymbol{z}$}{General variable vector}
\nomenclature{$\alpha$}{Strong convexity parameter}
Based on the distance-generating function $\omega(\cdot)$, we then introduce a prox-function (also known as the Bregman divergence \cite{Bregman1967}) $B: X \times X \rightarrow \mathbb{R}_{+} $ as:
\begin{equation}\label{eq:prox}
B(\boldsymbol{x},\boldsymbol{z})=
\omega(\boldsymbol{z})-\left[\omega(\boldsymbol{x})+
\nabla\omega(\boldsymbol{x})^T(\boldsymbol{z}-\boldsymbol{x})\right]
\end{equation}
\nomenclature{$V\left(\cdot,\cdot\right)$}{Prox function (Bregman distance)}
Notice that, due to the convexity of $\omega(\cdot)$, we can show that $B(\bm{x},\cdot)$ is non-negative. Associated with the distance-generating function $\omega(\cdot)$, a prox-mapping $P_{\boldsymbol{x}}: \mathbb{R}^n \rightarrow X$ can be defined as:
\begin{equation}\label{eq:prox-map}
P_{\boldsymbol{x}}(\boldsymbol{y})=\arg \min_{\boldsymbol{z} \in X}\left\lbrace \boldsymbol{y}^T(\boldsymbol{z}-\boldsymbol{x})+
B(\boldsymbol{x},\boldsymbol{z})\right\rbrace
\end{equation}
\nomenclature{$P\left(\cdot\right)$}{Prox-mapping}
\nomenclature{$\boldsymbol{y}$}{General variable vector}
Notice that because of the strong convexity of $B(\bm{x},\cdot)$, the above prox-mapping is well defined and has a unique value.
Making use of the prox-mapping, the MDSA update the design variable according to the following formula,
\begin{equation}\label{eq:MDSA update 1}
\boldsymbol{x}_{k+1} = P_{\boldsymbol{x}_k}\Big(\eta_k {\boldsymbol{G}}_m(\bm{x}_k)\Big)= \arg \min_{\boldsymbol{z} \in X}\Bigg\{ \Big({\boldsymbol{G}}_m(\bm{x}_k)\Big)^T(\boldsymbol{z}-\boldsymbol{x}_k)+ \frac{1}{\eta_k}B(\boldsymbol{x}_k,\boldsymbol{z})\Bigg\},
\end{equation}
\nomenclature{$\eta_k$}{Step size at $k$th step}
\nomenclature{$k$}{Step count of optimization step}
where $\bm{x}_k$, $\eta_k>0$, and ${\boldsymbol{G}}_m(\bm{x}_k)$ are the design variable, step size, and unbiased gradient estimator (using $m$ samples) at optimization step $k$, respectively. 

To gain a better understanding, let us take a closer look at the above update formula. The first term in the right bracket of \eqref{eq:MDSA update 1} is a linear approximation of the objective function at $\boldsymbol{x}_k$ using the gradient estimator ${\boldsymbol{G}}_m(\bm{x}_k)$, and the second term is a strongly convex function scaled by $1/\eta_k$. We can show that, at $\boldsymbol{z} = \boldsymbol{x}_k$, the gradient of $B(\boldsymbol{x}_k,\boldsymbol{z})$ vanishes and, as a result, the gradient of the entire expression in the bracket with respect to $\bm{z}$ equals to ${\boldsymbol{G}}_m(\bm{x}_k)$. In addition, the Hessian of the expression in the bracket equals to $\nabla^2\omega(\bm{z})$ scaled by $1/\eta_k$. This indicates that the expression in the bracket can be deemed as a convex approximation of the original objective function using the stochastic gradient ${\boldsymbol{G}}_m(\bm{x}_k)$, and the local curvature of the expression can be controlled through $\eta_k$.

We conclude this subsection with several remarks on the MDSA framework. First, same as the classical SA methods, the MDSA method can work with highly noisy gradient estimators. To ensure the convergence of MDSA update when applied to general stochastic optimization problems, only a single sample is required with a properly chosen step size policy \cite{Nemirovski2009}. This is firstly demonstrated in topology optimization by \cite{Zhang2020} for a randomized formulation to optimize structures under many deterministic load cases and the reduction to one sample load case. Alternatively, one can evaluate the gradient estimator ${\boldsymbol{G}}_m(\bm{x}_k)$ using multiple samples \cite{Zhang2017} and integrate with a commonly use update scheme (e.g., Optimality Criteria). 
%\textbf{In this work, we find that the MDSA update works well with any sample sizes $m\geq 2$.} 
Second, as compared to the classical SA approaches, the MDSA framework allows for a more general setup mainly because of the general definition of distance generating-function $\omega(\cdot)$. In fact, if we choose distance-generating function $\omega(\bm{x})=1/2||x||^2$ with $||\cdot||$ being the Euclidean norm, the MDSA update becomes the classic SA (or equivalently SGD) method \cite{Robbins1951,Nemirovski2009}. As demonstrated theoretically and numerically in \cite{Nemirovski2009} for general stochastic optimization problems, by choosing a proper distance-generating function, MDSA can adapt the update to the geometry of the problem, which leads to better performance in accuracy and convergence. This advantage is exploited in \cite{Zhang2020} for a deterministic topology optimization problem. 
%Third, the update formula \eqref{eq:MDSA update 1} can also be adopted in the deterministic setting with ${\boldsymbol{G}}_m(\bm{x}_k)$ being replaced with the exact gradient of the objective function $\bm{g}(\bm{x}_k)$, which leads to the Mirror Descent method \cite{Nemirovski1983} for deterministic optimization problems.

%Plugging in the expression and multiplying the constant $1/\eta_k$, we obtain the MDSA update formula:

%\begin{equation}\label{eq:MDSA update 2}
%\boldsymbol{x}_{k+1} = \arg \min_{\boldsymbol{z} \in X}\left\lbrace { \boldsymbol{G}\left(\boldsymbol{x}_k,\boldsymbol{\xi}_k\right)}^T(\boldsymbol{z}-\boldsymbol{x}_k)+ \frac{1}{\eta_k} V(\boldsymbol{x}_k,\boldsymbol{z})\right\rbrace.
%\end{equation

%As a special case, if we choose $\omega(\boldsymbol{x})=\frac{1}{2} \|\boldsymbol{x}\|_2^2$, the update \eqref{eq:MDSA update 1} becomes the stochastic projected gradient descent update \cite{Nemirovski2009}.

%The update formula \eqref{eq:MDSA update 1} reveals perhaps the biggest advantage of MDSA: \textit {at each optimization step, MDSA samples only once, reducing the step-wise cost to the level of deterministic updates!} Moreover, it is shown that by appropriately choosing the step size (called step size policy), the iterative process \eqref{eq:MDSA update 1} converges to the minimum of \eqref{eq:SA_Obj} \cite{Nemirovski2009}. We refer the readers to \cite{Nemirovski2009} for more technical details and convergence analysis.

\subsection{Accelerated Mirror Descent Stochastic Approximation (AC-MDSA)}
In general, the performance of the MDSA is sensitive to the choice of the step size policy. Large step size in MDSA could potentially lead to divergence, while too small step size may result in slow convergence. 
%This is also the case in our numerical examples. 
To alleviate this sensitivity, we introduce an accelerated version of MDSA \cite{Lan2012}, referred to as AC-MDSA, which makes use of momentum-based acceleration techniques. The AC-MDSA algorithm is proposed in \cite{Lan2012} for general stochastic optimization problems and is shown to achieve the optimal convergence rate for convex problems. The general update algorithm is presented in Algorithm \ref{alg: AC-MDSA}. Compared to the classic MDSA, which updates the $\boldsymbol{x}_{k}$ sequence, the AC-MDSA includes updating two additional sequences, namely a ``middle variable" $\boldsymbol{x}_k^{md}$ and an ``aggregated variable" $\boldsymbol{x}_k^{ag}$ \cite{Lan2012}. 

% 
%Having introduced the idea and update formula of MDSA, we proceed to one of the popular accelerated versions of MDSA, namely, AC-MDSA. The AC-MDSA algorithm in this study is based on \cite{Lan2012}. The AC-MDSA includes updating two additional sequences, namely a ``middle variable" $\boldsymbol{x}_k^{md}$ and an ``aggregated variable" $\boldsymbol{x}_k^{ag}$ \cite{Lan2012}. We present the pseudo-code first and provide the interpretations about the algorithm. The pseudo-code for the AC-MDSA algorithm is:

\begin{algorithm}[H]
	\caption{AC-MDSA}\label{alg: AC-MDSA}
	\begin{algorithmic}[1]
		\State \textbf{Initialize:} $\boldsymbol{x}_1$, $\boldsymbol{x}_1^{ag} = \boldsymbol{x}_1$, step sizes $\beta_1$ and $\eta_1$.
		
		\State \textbf{Set:} $\boldsymbol{x}_k ^{md} = \beta_k^{-1}\boldsymbol{x}_k + (1- \beta_k^{-1}) \boldsymbol{x}_k^{ag}$, with $\beta_k$ computed by \eqref{eq:stepsize_policy_beta}
		
		%		\State \textbf{Compute the sample gradient at $\boldsymbol{x}_k ^{md}$:} $ \boldsymbol{G}\left(\boldsymbol{x}_k ^{md},\boldsymbol{\xi}_k\right)$
		
		\State \textbf{Get $\boldsymbol{x}_{k+1}$ using MDSA update \eqref{eq:MDSA update 1}:} $\boldsymbol{x}_{k+1} = P_{\boldsymbol{x}_k} \left(\eta_k  {\boldsymbol{G}}_{m}\left(\boldsymbol{x}_k ^{md}\right) \right)$, with $\eta_k$ computed by \eqref{eq:stepsize_policy_eta}
		
		\State \textbf{Set:} $\boldsymbol{x}_{k+1}^{ag} = \beta_k^{-1}\boldsymbol{x}_{k+1} + \left(1 - \beta_k^{-1}\right)  \boldsymbol{x}_{k}^{ag}$ \label{eq:beta}
		
	\end{algorithmic}
\end{algorithm}
\nomenclature{$\boldsymbol{x}^{ag}$}{Aggregated variable vector in the AC-MDSA algorithm}
\nomenclature{$\boldsymbol{x}^{md}$}{Intermediate variable vector in the AC-MDSA algorithm}
\nomenclature{$\beta$}{Weighting coefficient to compute $\bm{x}_k^{md}$ and $\bm{x}_k^{ag}$ in the AC-MDSA algorithm}
The modifications from the standard MDSA update \eqref{eq:MDSA update 1} in the AC-MDSA algorithm mainly lie in three aspects. First, in addition to the sequence of $\boldsymbol{x}_k$, the algorithm updates the sequences $\boldsymbol{x}_k^{md}$ and $\boldsymbol{x}_k^{ag}$. 
We note that the converged sequence $\boldsymbol{x}_k^{ag}$ represents the final solution. As we show in the next subsection, $\boldsymbol{x}_N^{ag}$ represents the history weighted average of $\boldsymbol{x}_k$ from step 1 to step $N$ with a linear weight \cite{Lan2012}. The introduction of two additional sequences adds a negligible computational cost as they are vector additions. Second, the AC-MDSA algorithm performs update $\boldsymbol{x}_{k+1}$ using the gradient estimator evaluated at $\boldsymbol{x}_k^{md}$ instead of $\boldsymbol{x}_k$. Third, compared with MDSA, AC-MDSA requires the specification of $\beta_k$, which acts as a weight factor to compute $\bm{x}_k^{md}$ and $\bm{x}_k^{ag}$.

%-- moved to next subsection. When tailoring AC-MDSA algorithm to RTO, we find that, compared with the MDSA method, the AC-MDSA method can lead to accelerated convergence performance without an increase in computational cost. We also observe that the AC-MDSA method is considerably less sensitive with respect to the choice of step sizes, which allows us to use larger step sizes. These advantages will be demonstrated in the numerical examples in Section 5.

\subsection{An Entropic AC-MDSA tailored for robust topology optimization}
Having presented the general frameworks of MDSA and AC-MDSA, we now derive an AC-MDSA algorithm with the $\ell_1$-norm tailored for the RTO problem \eqref{eq:RTO original form.} and propose the explicit update formula. 
With the volume (linear) and box constraints, the feasible set $X$ of the RTO formulation \eqref{eq:RTO original form.} is give by 
\begin{equation}
X = \lbrace \boldsymbol{x \in \mathbb{R}^n}: \frac{ V\left(\boldsymbol{x} \right)}{V_0} - V_f \leq 0, x^{(i)} \in \left[0,1\right],\quad i=1,...,n\rbrace
\end{equation}
We define a scaled design variable vector $\tilde{\bm{x}}$ such that $\tilde{x}^{(i)}: = \tilde{v}^{(i)} x^{(i)}$ with $\tilde{v}^{(i)}$ being $
\tilde{v}^{(i)} := { \left(\boldsymbol{H}^T \boldsymbol{v}\right)^{(i)}}/({V_0 V_f})$.
%\begin{equation}
%\tilde{v}^{(i)} := \frac{ \left(\boldsymbol{H}^T \boldsymbol{v}\right)^{(i)}}{V_0 V_f},
%\end{equation}
The corresponding feasible set of the scaled variable is: 
\begin{equation}
\tilde{X} = \lbrace \boldsymbol{\tilde{x} \in \mathbb{R}^n}: \sum_{i=1}^{n} \tilde{x} ^{(i)} - 1 \leq 0, x^{(i)} \in \left[0,\tilde{v}^{(i)}\right],\quad i=1,...,n\rbrace
\end{equation}
Accordingly, the gradient estimator of the objective function in \eqref{eq:RTO original form.} with respect to $\tilde{\bm{x}}$ is obtained as:
%\begin{equation}
%\nabla_{\tilde{\boldsymbol{x}}} C(\tilde{\boldsymbol{x}},\boldsymbol{\xi}_j) = \text{diag}\left(\frac{1}{\tilde{v}^{(i)}}\right)\nabla_{\boldsymbol{x}} C(\boldsymbol{x},\boldsymbol{\xi}_j) 
%\end{equation}
%
\begin{equation} \label{eq: sensitivity chain rule}
{\tilde{\boldsymbol{G}}}_m \left(\tilde{\boldsymbol{x}}\right) = \text{diag}\left(\frac{1}{\tilde{v}^{(i)}}\right) {\boldsymbol{G}}_m \left(\boldsymbol{x}\right)
\end{equation}
\nomenclature{${\tilde{\boldsymbol{G}}}_m$}{Stochastic gradient with respect to scaled design variables estimated with $m$ samples}
%where the $i$th entry of ${\boldsymbol{\tilde{G}}}_m \left(\tilde{\boldsymbol{x}}\right)$ is $\partial{{J}_m \left(\boldsymbol{x}\right)} / \partial{\tilde{x}^{(i)}}$.

The feasible set $\tilde{X}$ is similar to a standard simplex set. Thus, we choose the $\ell_1$-norm with entropy function as $\omega$ in the proposed AC-MDSA algorithm, denoted as entropic AC-MDSA, because this setup leads to an improved convergence performance over the $\ell_2$-norm setting (which leads to the classical SA/SGD method) for a simplex set \cite{Zhang2020}. 
%Therefore, the AC-MDSA algorithm proposed in this work is based upon a type of MDSA with the $\ell_1$-norm setting and entropy function being its distance-generating function. 
%For now on, we will refer to this type of MDSA the . 
%In the sequel, we will first derive the detailed update formula for entropic MDSA. The derived formula is then used in the acceleration Algorithm \ref{alg: AC-MDSA} to obtain the proposed entropic AC-MDSA.
The distance-generating function $\omega$ of the entropic AC-MDSA takes the following form,
\begin{equation}\label{eq:entropy func}
\omega(\tilde{\boldsymbol{x}}) = \sum_{i=1}^{n} \tilde{{x}}^{(i)} \ln \tilde{{x}}^{(i)},
\end{equation}
and the corresponding prox-function becomes
\begin{equation}\label{eq:prox entropy}
B(\tilde{\boldsymbol{x}},\boldsymbol{z})=
\sum_{i=1}^{n} \left( z^{(i)} \ln{\frac{z^{(i)}}{\tilde{{x}}^{(i)}} - z^{(i)} + \tilde{{x}}^{(i)}} \right)
\end{equation}
\nomenclature{$z^{(i)}$}{General variable associated with element $i$}
By plugging \eqref{eq:prox entropy} into the prox-mapping \eqref{eq:prox-map} and dropping the constant terms, the update formula \eqref{eq:MDSA update 1} becomes
\begin{equation}\label{eq:update entropy}
\tilde{\boldsymbol{x}}_{k+1} =
P_{\tilde{\boldsymbol{x}}_k} \left(\eta_k  {\tilde{\boldsymbol{G}}}_m\left(\tilde{\boldsymbol{x}}_k\right) \right)
=\arg \min_{\boldsymbol{z} \in \tilde{X}}\left\lbrace { \Big({\tilde{\boldsymbol{G}}}_m\left(\tilde{\boldsymbol{x}}_k\right)}\Big)^T \boldsymbol{z}+ \frac{1}{\eta_k} \sum_{i=1}^{n} \left( z_i \ln{\frac{z^{(i)}}{\tilde{{x}}_k^{(i)}} - z^{(i)}} \right)\right\rbrace
\end{equation}
The above formula is given as a minimization problem, where $\tilde{\boldsymbol{x}}_{k+1}$ is its unique minimizer. Next, we derive an explicit update formula for \eqref{eq:update entropy}. 
The Lagrangian of \eqref{eq:update entropy} with respect to the (scaled) volume constraint is given by
\begin{equation}\label{eq:Lagrangian}
L(\boldsymbol{z},\lambda) =
{\tilde{\boldsymbol{G}}}_m^T \left(\tilde{\boldsymbol{x}}_k\right) \boldsymbol{z}+ \frac{1}{\eta_k} \sum_{i=1}^{n} \left( z^{(i)} \ln{\frac{z^{(i)}}{\tilde{x}_k^{(i)}} - z^{(i)}} \right) + \lambda (\boldsymbol{l}^T \boldsymbol{z} - 1), 
\end{equation}
\nomenclature{$\lambda$}{Lagrange multiplier}
\nomenclature{$L\left(\cdot,\cdot\right)$}{Lagrangian}
where $\lambda\in\mathbb{R}_{+}$ is the Lagrange multiplier associated with the constraint and $\boldsymbol{l}$ is a constant vector whose components are all 1. Imposing the gradient condition $\nabla_{\boldsymbol{z}} L(\boldsymbol{z},\lambda) = \boldsymbol{0}$ gives:
\begin{equation}\label{eq:grad Lagrangian}
\frac{\partial L(\boldsymbol{z},\lambda)}{\partial z_i} =
{\tilde{G}}_m^{(i)} \left(\tilde{\boldsymbol{x}}_k\right) + \frac{1}{\eta_k} \ln \frac{z^{(i)}}{\tilde{x}_k^{(i)}} + \lambda = 0,
\end{equation}
which can be recast as
\begin{equation}\label{eq:entropic MDSA update}
z^{(i)} \left(\lambda \right) = \tilde{x}_k^{(i)} \exp\left(-\eta_k \left({\tilde{G}}_m^{(i)} \left(\tilde{\boldsymbol{x}}_k\right) + \lambda\right)\right)
\end{equation}
Incorporating the box constraints, we then obtain the update formula:
\begin{equation}\label{eq:entropic MDSA update 2}
\tilde{x}_{k+1}^{(i)} (\lambda^{*}) = \begin{cases}
& z^{(i)} \left(\lambda^{*} \right), 	\quad \text{if} \quad \underbar{$\tilde{x}$}_{k+1}^{(i)} \leq z^{(i)} \left(\lambda^{*}\right) \leq \bar{\tilde{x}}_{k+1}^{(i)}, \\
& \bar{\tilde{x}}_{k+1}^{(i)},		\quad \quad \text{if} \quad z^{(i)} \left(\lambda^{*} \right) > \bar{\tilde{x}}_{k+1}^{(i)}, \\
& \underbar{$\tilde{x}$}_{k+1}^{(i)},	\quad \quad\text{if} \quad z^{(i)} \left(\lambda^{*} \right) < \underbar{$\tilde{x}$}_{k+1}^{(i)},
\end{cases}
\end{equation}

\nomenclature{$move$}{Move limits for updating design variables}
\nomenclature{$\bar{\tilde{x}}$}{Upper bound of updated design variables}
\nomenclature{$\underbar{$\tilde{x}$}$}{Lower bound of updated design variables}
\noindent
where $\bar{\tilde{x}}_{k+1}^{(i)} \doteq \text{min} \{ \tilde{x}_k^{(i)} + \tilde{v}^{(i)} move, \tilde{v}^{(i)}\}$ is the upper bound of updated design variables and $\underbar{$\tilde{x}$}_{k+1}^{(i)} \doteq \text{max} \{ \tilde{x}_k^{(i)} - \tilde{v}^{(i)} move, 0\}$ is the lower bound, and $\lambda^{*}$ solves equation $\sum_{i=1}^{n}\tilde{x}^{(i)}_{k+1} (\lambda) = 1$ \footnote{Because compliance problems have active volume constraints in practice, the proposed update formula assumes that mapped volume constraint is active throughout the optimization process, namely, $\sum_{i=1}^{n}\tilde{x}^{(i)}=1$.}. In practice, $\lambda^{*}$ is obtained using the bi-section method. Notice that, in \eqref{eq:entropic MDSA update 2}, we introduce a move limit denoted as $move$, which will be adaptively adjusted by a damping scheme (see Section \ref{section:damping}). After obtaining $\tilde{\boldsymbol{x}}_{k+1}$, we map it back to the original feasible space as
\begin{equation} \label{eq: back scaling}
x_{k+1}^{(i)}: = \frac{1}{\tilde{v}^{(i)}} \tilde{x}_{k+1}^{(i)}
\end{equation}
Finally, the proposed entropic AC-MDSA for RTO problem \eqref{eq:RTO original form.} is obtained by replacing \eqref{eq:MDSA update 1} in the step $3$ of Algorithm \eqref{alg: AC-MDSA} with \eqref{eq:entropic MDSA update}, \eqref{eq:entropic MDSA update 2}, and \eqref{eq: back scaling}. 
%
%\begin{algorithm}[H]
%	\caption{Entropic AC-MDSA for Robust Topology Optimization}\label{alg: AC-MDSA-entropic}
%	\begin{algorithmic}[1]
%		\State \textbf{Initialize:} $\boldsymbol{x}_0$, $\boldsymbol{x}_0^{ag} = \boldsymbol{x}_0$, step sizes $\beta_0$ and $\eta_0$.
%		\State Define $\tilde{\boldsymbol{x}}_0 = \text{diag}\left({\tilde{v}^{(i)}}\right) \boldsymbol{x}_0$,
%		\State \textbf{Set:} $\boldsymbol{x}_k ^{md} = \beta_k^{-1}\boldsymbol{x}_k + (1- \beta_k^{-1}) \boldsymbol{x}_k^{ag}$
%		
%		\State \textbf{Get $\boldsymbol{x}_{k+1}$ using update \eqref{eq:entropic MDSA update 2} and scaling \eqref{eq: back scaling}} 
%		
%		\State \textbf{Set:} $\boldsymbol{x}_{k+1}^{ag} = \beta_k^{-1}\boldsymbol{x}_{k+1} + \left(1 - \beta_k^{-1} \boldsymbol{x}_{k}^{ag}\right) $
%		
%	\end{algorithmic}
%\end{algorithm}
Several remarks can be made regarding the above entropic AC-MDSA update. First, the derived update \eqref{eq:entropic MDSA update} and \eqref{eq:entropic MDSA update 2} can handle both positive and negative stochastic gradients, thus it is also applicable to RTO problems with other objective functions, such as the compliance mechanism design \cite{Kogiso2008,Cardoso2019}. Second, as long as we start from a feasible initial guess, the $\tilde{\boldsymbol{x}}_k$ (and $\boldsymbol{x}_k$) always stays positive. Thus, the lower bound $x_{min}=0$ of the design variables is typically not active.
By tailoring AC-MDSA algorithm to RTO, we find that, compared with the MDSA method, the AC-MDSA method can lead to accelerated convergence performance without an increase in computational cost. We also observe that the AC-MDSA method is considerably less sensitive to the choice of step sizes, which allows us to use larger step sizes. These advantages will be demonstrated in the numerical examples in Section \ref{section:example}.

%%%%%%%%%%%%%%%%%%%%%%%%%%%%%%%%%%%%%%%%%% END of Section 3 %%%%%%%%%%%%%%%%%%%%%%%%%%%%%%%%%%%%%%%%%%%%%%%%%%

\section{Algorithmic parameters and implementation details of the entropic AC-MDSA for RTO}

This section discusses the algorithmic and implementation details of the proposed entropic AC-MDSA algorithm for RTO problems. In particular, we present the step size policy and introduce related techniques (i.e., step size recalibration and adaptive damping) to accelerate the convergence performance and reduce the step-size sensitivity of the AC-MDSA algorithm. 

%\subsection{Algorithm of solving the RTO problem with the two-sample entropic AC-MDSA }

\subsection{Step size policy}

Typically, the step size policy is critical for stochastic optimization algorithms. The step size policy adopted in this work is based on \cite{Lan2012} and involves two sequences $\eta_k$ and $\beta_k$.
The policy for $\eta_k$ is given by
\begin{equation} \label{eq:stepsize_policy_eta}
\eta_{k} = \theta \,\bar{\eta}\, \frac{k+1}{2},\quad k=1,2,...,N,
\end{equation}
where $\theta$ is a user-defined scaling factor that adjusts the step size, and $\bar{\eta}$ is computed according to
\begin{equation} \label{eq: step size rule 2}
\bar{\eta} = \frac{\sqrt{6 \alpha} D_{\omega, \tilde{X}}}{(N+2)^{\frac{3}{2}}\left(4 \mathcal{M}^{2}+\Sigma^{2}\right)^{\frac{1}{2}}},
\end{equation}

\nomenclature{$\theta$}{Step size scaling factor of AC-MDSA algorithm}
\nomenclature{$D_{\omega, \tilde{X}}$}{Diameter of set $\tilde{X}$ defined by distance generating function $\omega$}
\nomenclature{$N$}{Maximum number of optimization steps}
\nomenclature{${\mathcal{M}}$}{Sample-based estimate of the upper bound for $\ell_\infty$-norm of the stochastic gradient mean}
\nomenclature{${\Sigma}$}{Sample-based estimate of the upper bound of $\ell_\infty$-norm of the stochastic gradient variance}
\nomenclature{$t$}{Summation count}
\nomenclature{$N_M$ {and} $N_{\Sigma}$}{Numbers of evaluations to estimate the upper bounds of the stochastic gradients and its variance in the step size policy, respectively}
\noindent
where $\alpha=1$, $D_{\omega,\tilde{X}}$ is the diameter of set $\tilde{X}$ measured by the distance-generating function $\omega$, which is taken to be $\sqrt{\ln(n)}$ \cite{Lan2012,Nemirovski2009}. Parameters $\mathcal{M}$ and $\Sigma$ are estimates of the upper bounds of the stochastic gradients and its variance. In this work, they are estimated using sampling-based techniques,
\begin{equation} \label{eq: sample-based M}
\mathcal{M} =\sqrt{ \frac{1}{N_M} \sum_{i=1}^{N_M} \left\| {\tilde{\boldsymbol{G}}}^{(i)}_m\left(\tilde{\boldsymbol{x}}_k\right)\right\|_\infty^2} \quad \text{and}\quad \Sigma = \sqrt{\frac{1}{N_{\Sigma}} \sum_{i=1}^{N_{\Sigma}} \left\| {\tilde{\boldsymbol{G}}}^{(i)}_m\left(\tilde{\boldsymbol{x}}_k\right) - \boldsymbol{\mathcal{Q}} \right\|_\infty^2},
\end{equation}
where ${\tilde{\boldsymbol{G}}}^{(i)}_m$ denotes the $i$th evaluation of ${\tilde{\boldsymbol{G}}}_m$ using an independent set of $m$ samples, $N_M$ {and} $N_{\Sigma}$ are the numbers of evaluations to estimate the upper bounds of the stochastic gradients and its variance, respectively, and $\boldsymbol{\mathcal{Q}} = {1}/{N_M} \sum_{i=1}^{N_M}  {\tilde{\boldsymbol{G}}}^{(i)}_m\left(\tilde{\boldsymbol{x}}_k\right)$.
%
%We use $N_M = N_{\Sigma} = 6$ to estimate $\mathcal{M}$ and $\Sigma$. 
%Since each evaluation of ${\tilde{\boldsymbol{G}}}_m$ requires $m$ samples, the total number of sampling is $6m$. 
%We note that, every time we need to compute/re-evaluate $\eta$ in our implementation, $\mathcal{M}$ and $\Sigma$ are estimated based on the above expression using the current design variables $\tilde{\boldsymbol{x}}_k$.
%
The policy for $\beta_k$ is given by 
\begin{equation}\label{eq:stepsize_policy_beta}
\quad\beta_{k} = \frac{k+1}{2},\quad k=1,2,..., N
\end{equation}
%\begin{equation}
%\begin{aligned}
%\mathcal{M}^2=\sup_{\tilde{\bm{x}}\in\tilde{X}} 
%\E \left[\left\| {\boldsymbol{G}}_m\left(\tilde{\boldsymbol{x}}\right)\right\|_\infty^2\right],
%\mathbb{E}\left[\left\| \boldsymbol{G}\left(\tilde{\boldsymbol{x}},\boldsymbol{\xi}\right)-\boldsymbol{g}\left(\tilde{\boldsymbol{x}}\right)\right\|_{\infty}^{2}\right] \leq \sigma^{2} \quad \forall \tilde{\boldsymbol{x}} \in \tilde{X}.
%\end{aligned}
%\end{equation}
%
%In the RTO problem, the two suprema are inaccessible, and therefore, we use a sample-based estimate ${\mathcal{M}}$ and ${\sigma}$ at the initial step:
%\begin{equation} \label{eq: sample-based M}
%{\mathcal{M}}^2 = \frac{1}{m} \sum_{i=1}^{m} \left\| \boldsymbol{G}\left(\tilde{\boldsymbol{x}}_0,\boldsymbol{\xi}_i\right)\right\|_\infty^2,
%\end{equation}
%%
%\begin{equation} \label{eq: sample-based sigma}
%{\sigma}^2 = \frac{1}{m} \sum_{i=1}^{m} \left\| \boldsymbol{G}\left(\tilde{\boldsymbol{x}}_0,\boldsymbol{\xi}_i\right) - {\mathcal{M}} \right\|_\infty^2.
%\end{equation}
%
If we plug \eqref{eq:stepsize_policy_beta} into the expression of aggregated variable $\tilde{\boldsymbol{x}}_{k+1}^{ag}$ in Algorithm \eqref{alg: AC-MDSA}, $\tilde{\boldsymbol{x}}_{k+1}^{ag}$ can be recast as \cite{Lan2012}:
\begin{equation}
\tilde{\boldsymbol{x}}_{k+1}^{ag} = \frac{2}{k+1} \tilde{\boldsymbol{x}}_{k+1} + \frac{k-1}{k+1} \tilde{\boldsymbol{x}}_k^{ag} = \frac{\sum_{t=1}^{k}\left(t \tilde{\boldsymbol{x}}_{t+1}\right)}{\sum_{t=1}^{k} t}
\end{equation}
The above expression indicates that the aggregated variable $\tilde{\boldsymbol{x}}_{k+1}^{ag}$, obtained by adopting $\beta$ policy \eqref{eq:stepsize_policy_beta}, is the weighted average of the history of variable $\tilde{\boldsymbol{x}}_{t+1}$ with linearly varying weights. We note that history-averaging techniques are widely used in SA methods to suppress noise and accelerate convergence \cite{Nemirovski2009,Polyak1992}. Such a noise-suppressing strategy is different from Monte Carlo based methods \cite{Rubinstein2016}, in which the noise is reduced through estimations using many samples within one optimization step.

\subsection{Adaptive step recalibration scheme}
The history-averaging technique can lead to a small change of designs as the optimization proceeds, leading to slow convergence. To address this, we propose a step size recalibration scheme to speed up the evolution of the design and convergence.

The basic idea of the recalibration scheme is to adaptively re-initialize the acceleration method throughout the optimization by tracking the changes of design variables. Specifically, we monitor the $\ell_2$-norm of the change of $\boldsymbol{x}_k^{ag}$, namely,
\begin{equation} \label{eq: recalibration}
\left\|\Delta \boldsymbol{x}_k^{ag}\right\|_2 = \left\|\boldsymbol{x}_{k+1}^{ag} - \boldsymbol{x}_k^{ag}\right\|_2
\end{equation}
If $\left\|\Delta \boldsymbol{x}_k^{ag}\right\|_2$ becomes smaller than a tolerance $\epsilon_{rst}$, then we recompute $\bar{\eta}$, $\mathcal{M}$, and $\Sigma$ based on \eqref{eq: step size rule 2} and \eqref{eq: sample-based M}, and set $k=1$ in evaluating $\eta_k$ and $\beta_k$. To avoid frequent recalibration, we require the number of optimization steps between two consecutive recalibrations to be larger than a prescribed minimum step $\Delta_{rst}$, and monitoring of $\left\|\Delta \boldsymbol{x}_k^{ag}\right\|_2$ starts after the first $N_{rst}$ steps.
We note that recalibration schemes of similar forms are shown to be effective for accelerated gradient descent methods in other applications, for example, see \cite{Zhang2020}.

\nomenclature{$N_{rst}$}{Starting step of step size recalibration monitoring}
\nomenclature{$\Delta_{rst}$}{Minimum interval steps of two consecutive step size recalibrations}

\subsection{Adaptive damping scheme}\label{section:damping}
Because of the stochastic nature of the entropic AC-MDSA algorithm, optimization terminates at the maximum step unless a decaying step size policy is adopted. Thus, this work adopts the adaptive damping scheme proposed in \cite{Zhang2017} to effectively terminate the optimization after the design has converged. Inspired by the simulated annealing \cite{Kirkpatrick1983,Salamon2002}, the adaptive damping scheme monitors the average progress of the design at each step and reduces the move limit when small progress is detected. The average progress of the design at $k$th step is characterized by the effective step ratio, $R_k$, which is defined as:
\begin{equation} \label{eq: step ratio}
R_k := \frac{\frac{1}{N_D} \left\|\bm{E}_k - \bm{E}_{k-N_D+1} \right\|_2 }{\left\|\bm{E} - \bm{E}_{k-1}\right\|_2},
\end{equation}
where $\bm{E}$ is the vector of elemental Young's moduli, $N_D$ is the history window size.

The effective step ratio, $R_k$, represents the relative magnitude of the average design change over the past $N_D$ steps to the current design change. A small $R_k$ indicates slow progress over the previous $N_D$ steps,
%POORLY WRITTEN: meaning the design has been wandering around rather than marching steadily in certain directions. 
the move limit is then reduced.
%to limit the random walk of the optimizer. 
%Based on this idea, the adaptive damping scheme works as follows: once
Specifically, if $R_k$ is lower than a tolerance $\epsilon_{damp}$, the move limit is scaled down by a factor $\tau$, i.e. $move = move/\tau$. Here, we use $\tau = 2$. 
%In addition, to prevent damping the update too early, 
The adaptive damping scheme is activated after a prescribed minimum number of steps $N_{damp}$.

\nomenclature{$R$}{Effective step ratio}
\nomenclature{$N_D$}{Window size of computing effective step ratio}
\nomenclature{$\tau$}{Reduction factor of move limits}
\noindent

%\subsection{Gradual decrease of filter radius}
%We use the gradual decrease of the filter radius to reduce gray areas in the final designs. As mentioned in the previous section, the physical density $\bar{\boldsymbol{x}}_k$ is the filtered version of the aggregated variable, i.e. $\bar{\boldsymbol{x}}_k = \boldsymbol{H}\boldsymbol{x}_k^{ag}$. Since the $\boldsymbol{x}_k^{ag}$ is the history average of $\boldsymbol{x}_k$ from all previous steps (since each restart), and $\boldsymbol{x}_k$ inherently involves randomness due to the random sampling, it is in general difficult to guarantee $100 \%$ 0-1 values in $\boldsymbol{x}_k^{ag}$. Therefore, we take additional measures to reduce the gray, and the gradual decrease of filter radius is quite effective.
%
%The decrease starts after a prescribed number of steps $k_{reduce}$ when the main structure has formed. This is often achieved after approximately $1/2$ to $2/3$ of the maximum iteration. The filter radius decreases by a fixed amount every certain steps until it reaches a preset minimum value:
%\begin{equation} \label{eq: decrease radius}
%r = \max \lbrace r - \Delta r, r_{min}\rbrace.
%\end{equation}
%\nomenclature{$r$}{Filter radius}
%\nomenclature{$r$}{Mimimal filter radius}

\subsection{Algorithm summary}
To conclude this section, we %summary
summarize the proposed entropic AC-MDSA algorithm and its parameters in Algorithm \ref{alg: AC-MDSA-RTO}. The objective function value and design quality are generally insensitive to most of the algorithm parameters, i.e., $N_{rst}$, $\Delta_{rst}$, $\epsilon_{rst}$, $N_{damp}$, $\epsilon_{damp}$,  $\tau$, $N_{D}$, $N_{max}$, $N_{min}$, and $\epsilon$. We have investigated various parameter values and summarized the value ranges used in this study in Table ~\ref{Tbl-Parameter values}, which are generally recommended. The step size scaling factor, $\theta$, has more influence on the results, as it directly adjusts the magnitude of the step size $\eta_k$. In general, a larger $\theta$ (and therefore larger $\eta_k$) leads to faster convergence and design evolution, but $\theta$ should not be too large as it may result in instability. The proper range of $\theta$ needs to be calibrated with a few pilot runs, but in general, the range that produces a stable and steady convergence is wide.

% \usepackage{multirow}
\begin{table}[H]
	\caption{Range of parameter values for AC-MDSA}
	\label{Tbl-Parameter values}
	\centering
	\begin{tabular}{ccc}
		\hline \hline
		Parameter         & Value & Usage                                        \\ \hline
		$N_{rst}$         & 100 or 300      & \multirow{3}{*}{Adaptive step recalibration} \\
		$\Delta_{rst}$    & 100       &                                              \\
		$\epsilon_{rst}$  &  0.025     &                                              \\ \hline
		$N_{damp}$        &    400 $\sim$  450   & \multirow{4}{*}{Adaptive damping scheme}     \\
		$\epsilon_{damp}$ & 0.05 or 0.075        &                                             \\
		$\tau$            &  2     &                                              \\
		$N_D$             & 100      &                                               \\ \hline
		$N_{max}$         & 450 $\sim$  600       &  \multirow{3}{*}{Termination of optimization} \\
		$N_{min}$         & 400 $\sim$  450      &                                              \\
		$\epsilon$        & 0.01  &                                              \\ \hline
		$N_M$        & 6  & \multirow{2}{*}{Estimation of $\mathcal{M}$ and $\Sigma$ \eqref{eq: sample-based M}}                                             \\
		$N_{\Sigma}$      & 6 \\ \hline
		
	\end{tabular}
\end{table}

\begin{algorithm}[H]
	\caption{Entropic AC-MDSA algorithm for robust topology optimization}\label{alg: AC-MDSA-RTO}
	\begin{algorithmic}[1]
		\State \textbf{Initialize:} $\boldsymbol{x}_1$, $\theta$ % $N_{rst}$, $N_{damp}$, $N_{min}$, $N_{max}$, $\tau$, $\epsilon_{rst}$, $\epsilon_{damp}$ and $\epsilon$.
		\State Set $\tilde{\boldsymbol{x}}_1 = \text{diag}\left({\tilde{v}^{(i)}}\right) \boldsymbol{x}_1$, $\tilde{\boldsymbol{x}}_1^{ag} = \tilde{\boldsymbol{x}}_1$; compute $\bar{\eta}$ using \eqref{eq: step size rule 2}; and set $k_{in} = 1$.
		\For{$k = 1, ..., N_{max}$}
		
		%		\If {$k\geq k_{reduce}$ and $\mod\left(k, int_{reduce}\right) = 0$}
		%			\State  Reduce filter radius using \eqref{eq: decrease radius} and reform filter matrix $\boldsymbol{H}$
		%		\EndIf		
		%		\State \textbf{end if}
		
		\If {$k\geq N_{rst}$ and $k_{in}\geq \Delta_{rst}$ and $|| \boldsymbol{x}_{k+1}^{ag}-\boldsymbol{x}_{k}^{ag}||_2 < \epsilon_{rst}$}
		\State  Set $k_{in} = 1$ and $\tilde{\boldsymbol{x}}_k = \tilde{\boldsymbol{x}}_k^{ag}$
		\State  Compute $\bar{\eta}$ using \eqref{eq: step size rule 2}
		\EndIf		
		\State \textbf{end if}
		\State Set $\tilde{\boldsymbol{x}}_k ^{md} = \beta_{k_{in}}^{-1}\tilde{\boldsymbol{x}}_k + (1- \beta_{k_{in}}^{-1}) \tilde{\boldsymbol{x}}_k^{ag}$ with $\beta_{k_{in}}$ defined in \eqref{eq:stepsize_policy_beta}.
		\State Compute gradient estimator ${\tilde{
				\bm{G}}}_m \left(\tilde{\boldsymbol{x}}_k ^{md}\right)$ according to \eqref{eq: sensitivity chain rule} and \eqref{eq: sensitivity} using $m$ i.i.d samples.
		
		\State Update $\tilde{\boldsymbol{x}}_{k+1}$ using entropic MDSA \eqref{eq:entropic MDSA update}--\eqref{eq:entropic MDSA update 2}
		
		\State Set $\tilde{\boldsymbol{x}}_{k+1}^{ag} = \beta_{k_{in}}^{-1}\tilde{\boldsymbol{x}}_{k+1} + \left(1 - \beta_{k_{in}}^{-1}\right) \tilde{\boldsymbol{x}}_{k}^{ag}$ with $\beta_{k_{in}}$ defined in \eqref{eq:stepsize_policy_beta}.
		
		\State Compute $\boldsymbol{x}_{k+1}^{ag}$ using \eqref{eq: back scaling}
		
		\If {$k\geq N_{min}$ and $||\boldsymbol{x}_{k+1}^{ag}-\boldsymbol{x}_{k}^{ag}||_{\infty} < \epsilon$}
		\State  break
		\EndIf		
		\State \textbf{end if}
		
		\State Evaluate effective step ratio $R_k$ using \eqref{eq: step ratio}
		
		\If {$k\geq N_{damp}$ and $R_k \leq \epsilon_{damp}$}
		\State  $move = move/\tau$
		\EndIf		
		\State \textbf{end if}
		\State  $k_{in}$ = $k_{in} + 1$
		\EndFor
		\State \textbf{end for}
		
		\State \textbf{Output:} $\boldsymbol{x}^{*} =\boldsymbol{x}_{k}^{ag}$
	\end{algorithmic}
\end{algorithm}

\nomenclature{$k_{in}$}{Counter of steps between two step size recalibrations}

%The We use the step-wise maximum change of the aggregated density $\|\Delta \boldsymbol{x}_k^{ag}\|_{\infty}$ as the indicator of convergence since $\boldsymbol{x}_k^{ag}$ represents the physical design, and a small value of $\|\Delta \boldsymbol{x}_k^{ag}\|_{\infty}$ means that the physical design has stabilized and converged. The stopping criteria is:
%\begin{equation} \label{eq: stopping criteria}
%\|\Delta \boldsymbol{x}_k^{ag}\|_{\infty} = \|\boldsymbol{x}_{k+1}^{ag} - \boldsymbol{x}_{k}^{ag} \|_{\infty} < \epsilon,
%\end{equation}
%where $\epsilon$ is the prescribed tolerance.

%%%%%%%%%%%%%%%%%%%%%%%%%%%%%% End of Section 4 %%%%%%%%%%%%%%%%%%%%%%%%%%%%%%%%%%%%%%%%%%%%%%%

\section{Numerical examples}\label{section:example}

This section presents four examples to demonstrate the effectiveness and efficiency of the entropic AC-MDSA algorithm. First, to verify the results by AC-MDSA, we compare the final designs, objective function values, and computational cost of the AC-MDSA with those from the Monte Carlo (MC) method. The MC method evaluates the sensitivity using $m=1,000$ samples at each optimization step to get sufficiently accurate gradients and uses a popular optimization update algorithm, MMA \cite{Svanberg1987}, to update the design variables with the estimated sensitivity. The second example shows that the AC-MDSA, although using two samples, effectively reflects the influence of $\kappa$ (relative weight of mean and variance) through both designs and objective function values. Example 3 demonstrates the AC-MDSA using problems with different domain geometries, multiple random loads, and various mesh sizes. Finally, in Example 4, we solve a three-dimensional (3D) problem to show the applicability of the entropic AC-MDSA with an iterative linear solver. The key information of the four examples is summarized in Table ~\ref{tbl-examples}. The investigated $\kappa$ values for the robust designs are $\kappa = 0.8284$, $0.618$, and $0.2824$, and they are chosen such that the equivalent ratio in terms of mean and standard deviation in the objective function with $w=1$, i.e. $\kappa : \sqrt{1-\kappa}$, is $2$, $1$, $\frac{1}{3}$, which are commonly used values in the RTO literature. The $\kappa$ values are summarized in Table ~\ref{Tbl-kappa values}.

\begin{table}[H]
	\caption{Investigated $\kappa$ values and their equivalent mean-to-s.t.d. ratios}
	\label{Tbl-kappa values}
	\centering
	\begin{tabular}{cc}
		\hline \hline
		$\kappa$ value & \begin{tabular}[c]{@{}c@{}}Equivalent ratio of mean : s.t.d.\\ ($\kappa : \sqrt{1-\kappa}$) with $w = 1$\end{tabular} \\ \hline
		1                           & -                                                                       \\
		0.828                      & 1 : 0.5                                                                                  \\
		0.618                       & 1 : 1                                                                                  \\
		0.282                      & 1 : 3                                                                                  \\ \hline
	\end{tabular}
\end{table}

We implement the proposed AC-MDSA algorithm in the PolyTop code \cite{Talischi2012}. To comprehensively and fairly evaluate the algorithm's performance, we carry out 50 consecutive and independent runs for each $\kappa$ studied in every 2D example and present the statistical data related to the algorithm's performance. Notice the 50 trials are only for evaluating statistical consistency and are not required for practical use of the algorithm. The presented design for each $\kappa$ is a representative design chosen from the 50 trials and has an objective function value close to the mean value of the 50 objective function values. 
% you didn't run 50 trials for the last example. Since the algorithm involves random sampling, we perform 50 consecutive and independent trials for each $\kappa$ studied in every example to get a fair and comprehensive evaluation of its performance. 
At the end of the optimization, denoting $\bm{x}^*$ as the optimized solution, we use $m=10,000$ samples to obtain accurate estimates of the objective function value, the mean, and the standard deviation of the compliance for the final design, denoted as $\hat{J}(\bm{x}^*)$, $\hat{\mu}(\bm{x}^*)$, and $\hat{\sigma}(\bm{x}^*)$, respectively. For comparison, we also include the deterministic designs with the objective function being the compliance under deterministic loads that take the mean values of the random loads. The $\hat{\mu}(\bm{x}^*)$ and $\hat{\sigma}(\bm{x}^*)$ of the optimized deterministic design is evaluated using the same random load corresponding to the stochastic cases. The total wall-clock time and the number of optimization steps are reported. 
%\textbf{CHECK THE SPECIFICATIONS!}
All the examples are performed on a machine with an Intel(R) Xeon(R) Silver 4116 CPU, 2.10GHz processor and 64 GB of RAM, running MATLAB R2018b.
In this work, the state equation is solved using the sparse direct solver and preconditioned conjugate gradient solver 
%\cite{MathWorks2019} 
for 2D and 3D problems, respectively. For most two-dimensional (2D) examples, we enforce the design symmetry about the vertical axis, and we study a 2D example without symmetry constraint. For the 3D example, we enforce design symmetry about the two vertical planes.

\begin{table}[H]
		\caption{Brief description of the numerical examples. }
			\label{tbl-examples}
	\begin{tabular}{lllll}
		\hline \hline
		Ex. & Dim. & Name                                                                  & Load uncertainty                                                                                                                      & Feature                                                                                                                                                                                                     \\ \hline
		1   & 2D   & \begin{tabular}[c]{@{}l@{}}Simple column\\ benchmark\end{tabular}     & \begin{tabular}[c]{@{}l@{}}Random direction\\ $\sim \mathcal{U}(\frac{11}{24}\pi,\frac{13}{24}\pi)$\end{tabular}                                                                          & \begin{tabular}[c]{@{}l@{}}- Verification of entropic\\    AC-MDSA with MC\\ - Study of sample size, \\    $m = 2, 10, 100$\\ - Comparison of MDSA algorithms\\    with and without acceleration\end{tabular} \\ \hline
		2   & 2D   & Half circle                                                           & \begin{tabular}[c]{@{}l@{}}Deterministic\\ vertical \& random\\ horizontal\\ components\\ $\sim \mathcal{N}(0,0.15^2)$ \end{tabular}                              & \begin{tabular}[c]{@{}l@{}}- Study of $\kappa$ values, \\    $kappa$ = 1, 0.618, 0.282\end{tabular}                                                                                                              \\ \hline
		3   & 2D   & \begin{tabular}[c]{@{}l@{}}Double hook\\ \& Torsion disk\end{tabular} & \begin{tabular}[c]{@{}l@{}}Deterministic\\ vertical/normal \&\\ multiple random\\ horizontal/tangential\\ components\\ $\sim \mathcal{N}(0,0.1^2)$\end{tabular} & \begin{tabular}[c]{@{}l@{}}- Problem size study, \\    $n$ = 114$k$, 51$k$, 13$k$\\ - Complex design geometries\\    and multiple independent \\    random components\\ -  Comparison with MC\end{tabular}          \\ \hline
		4   & 3D   & Crane                                                           & \begin{tabular}[c]{@{}l@{}}Deterministic z-direction,\\ multiple random x- \& y-\\ directions\\ $\sim \mathcal{N}(0,0.1^2)$\end{tabular}                        & \begin{tabular}[c]{@{}l@{}}- Combination of entropic \\    AC-MDSA with iterative\\ linear solver\end{tabular}                                                                                              \\ \hline
	\end{tabular}
\end{table}
\nomenclature{$\mathcal{N}(\cdot,\cdot)$}{Normal distribution}
\nomenclature{$\mathcal{U}(\cdot,\cdot)$}{Uniform distribution}
\subsection{Example 1: Simple column benchmark}
The first example is the simple column involving randomness in the load direction, which is commonly studied in the literature of RTO. We first verify the proposed entropic AC-MDSA (using two samples) by comparing its results with the ones obtained by the MC method (using 1000 samples). Then, we demonstrate the robustness of the entropic AC-MDSA algorithm with respect to different sample sizes $m$ (thus different accuracy levels) for computing the gradient estimator.
%, which leads to different accuracy levels of the gradient estimator. 
Finally, we compare the performance of the entropic AC-MDSA algorithm with the entropic MDSA (without acceleration).% and demonstrate that the entropic AC-MDSA produces more consistent designs in the absence of the symmetry constraint. 
%%less sensitive to step size choices than the entropic MDSA.

\begin{figure}[H]
	\centering
	\includegraphics[width=5in]{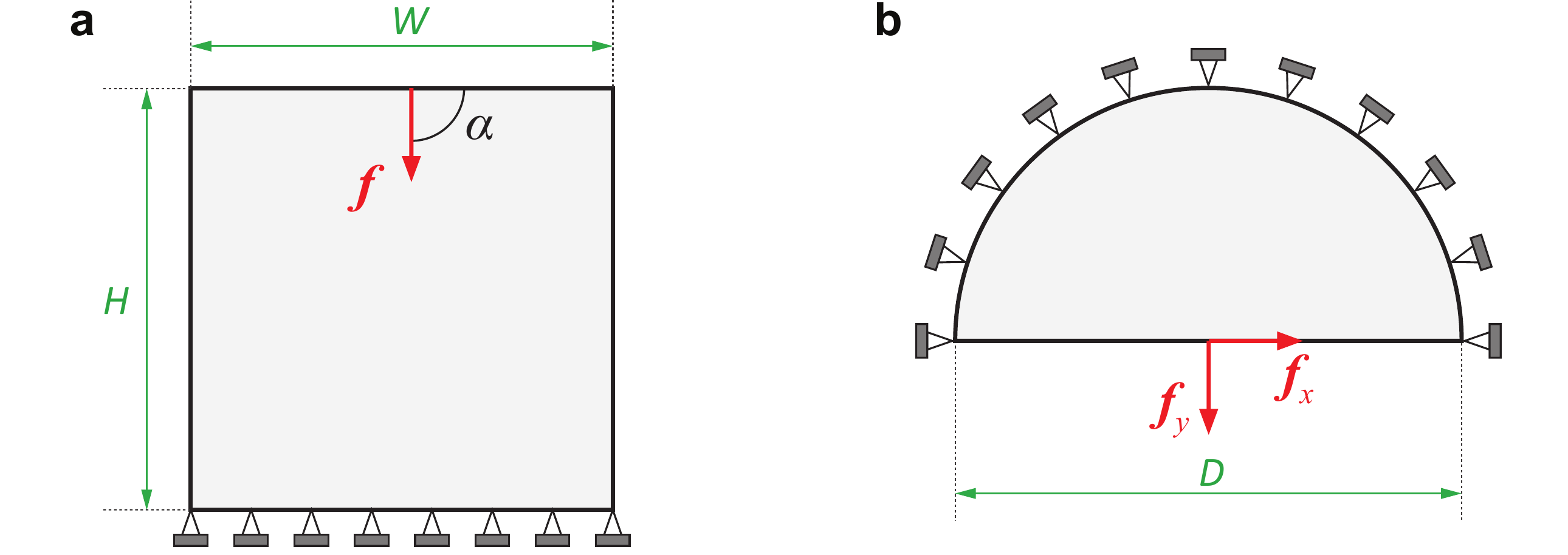}
	\caption{Geometry and boundary conditions of (a) Example 1: simple column, $H = W = 100$,  point load $\boldsymbol{f}$ has a deterministic magnitude of 1 and a random load direction $\alpha \sim \mathcal{U}(\frac{11}{24}\pi,\frac{13}{24}\pi)$; (b) Example 2: half circle, $D=1$, point load has a deterministic vertical component $\boldsymbol{f}_y = 1$ and a random horizontal component $\boldsymbol{f}_x \sim \mathcal{N}(0,0.15^2)$. }\label{Fig_Ex1_Ex2_geo}
\end{figure}

Figure \ref{Fig_Ex1_Ex2_geo}a shows the design domain and boundary conditions of the simple column problem. 
%The dimensions of the domain is . 
The domain is fixed at the bottom and is subjected to a load $\boldsymbol{f}$ with a deterministic magnitude of 1 and a random direction, defined by $\alpha \sim \mathcal{U}(\frac{11}{24}\pi,\frac{13}{24}\pi)$ with the standard deviation being $\frac{1}{12\sqrt{12}} \pi$, which is in the common range used in the literature \cite{Dunning2011,Zhao2014}. We consider three cases: a deterministic design ($\alpha \sim \mathcal{U}(\frac{1}{2}\pi,\frac{1}{2}\pi)$), a robust design with $\kappa = 1$, and a robust design with $\kappa = 0.618$.
The mesh size $n = 100 \times 100 = 10,000$, and the initial density filter radius is $R = 3$. For the entropic AC-MDSA algorithm, we use the sample size $m=2$, $\theta = 600n$, $N_{rst} = 100$, $N_{damp} = 400$, $\epsilon_{damp}=0.05$, $N_{max}=500$, and $N_{max}=400$. The filter radius begins to reduce to $R = 1.2$ with an interval of 30 steps. For the MC method, we use $m=1000$ and $N_{max}=100$, and the filter radius starts to decrease at the 60th step, which is at the same stage relative to the $N_{max}$ ($60/100 = 0.6$) as the one in AC-MDSA ($300/500 = 0.6$). We chose the relatively small $N_{max} = 100$ for the MC because the computational cost for MC with $m=1000$ samples is excessive.
%%
% Figure 1
\begin{figure}[H]
	\centering
	\includegraphics[width=6in]{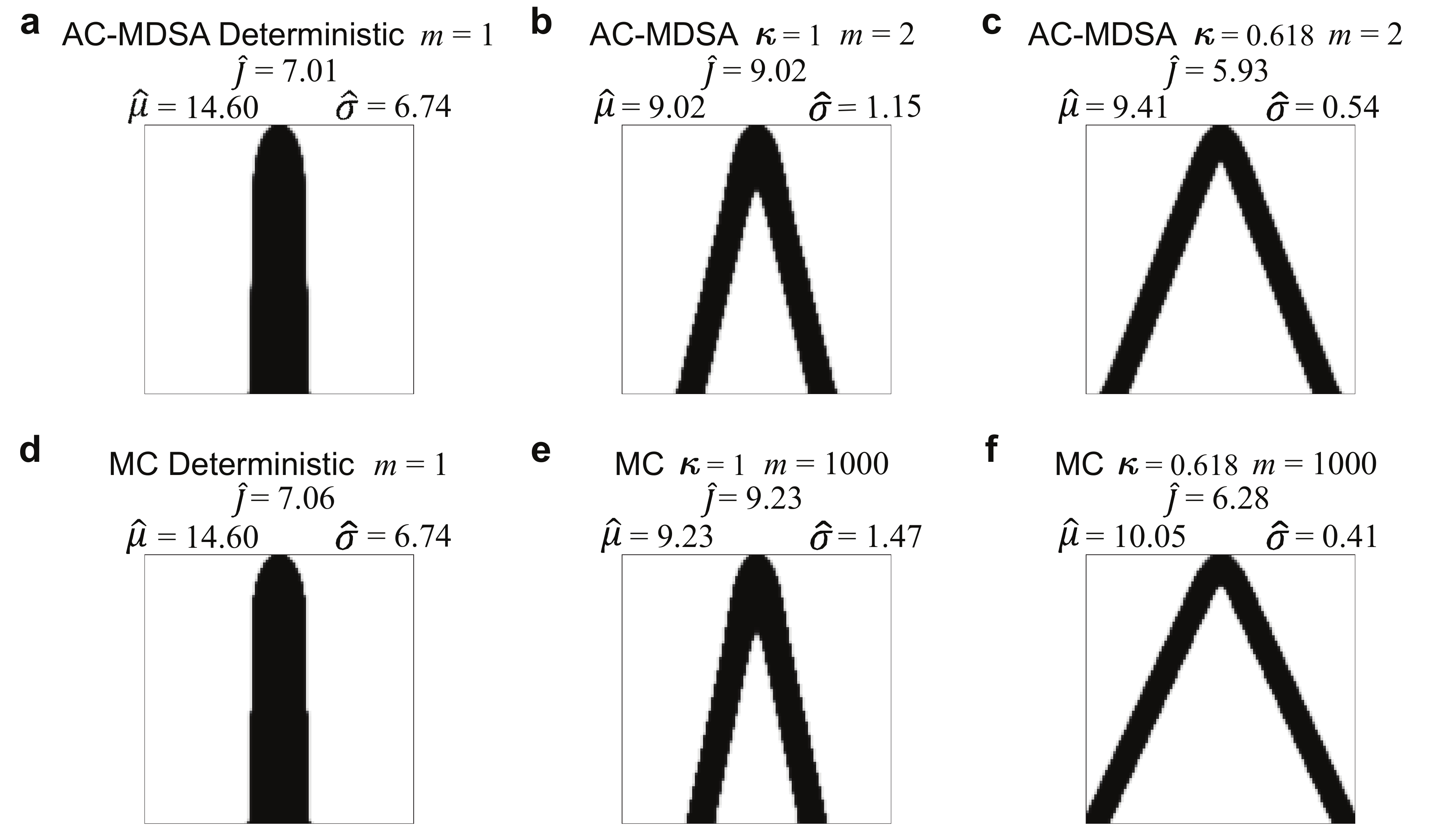}
	\caption{Final designs of (a) AC-MDSA, deterministic; (b) AC-MDSA, $\kappa=1$;  (c) AC-MDSA, $\kappa=0.618$;  (d) MC, deterministic; (e) MC, $\kappa=1$; (f) MC, $\kappa=0.618$. The design in (b) and (C), respectively, is a representative design chosen from the 50 trials.}\label{Fig_Ex1_Designs}
\end{figure}

\subsubsection{Verification of AC-MDSA with MC}
Here, we verify the entropic AC-MDSA algorithm with MC method by comparing the representative final designs and objective function values as shown in Figure \ref{Fig_Ex1_Designs} and statistics in Table ~\ref{Tbl-Ex_1}. For each $\kappa$, the representative design of the AC-MDSA in Figure \ref{Fig_Ex1_Designs} has an objective function value close to the mean of the objective function values of the 50 trials. 
%To evaluate the stability of the stochastic algorithm, for robust cases ($\kappa = 1$ and $\kappa = 0.618$) of AC-MDSA, we run the problem 50 times and present the statistics of the results in Table 1 (note that it is not needed in practice). The designs shown in Figure \ref{Fig_Ex1_Designs} are representative ones.
%selected from $50$ trials in that they have objective function values close to the mean values of the 50 trials, respectively. 
%
For the deterministic cases (Figures \ref{Fig_Ex1_Designs}a and \ref{Fig_Ex1_Designs}d), the entropic AC-MDSA and MC methods produce similar designs with comparable objective values, demonstrating that the entropic AC-MDSA can also be used to solve deterministic problems. Notice that in the deterministic case, even though the $\hat{\mu}$ and $\hat{\sigma}$ are identical for the AC-MDSA and MC, the $\hat{J}$ are different. This is because the $\hat{\mu}$ and $\hat{\sigma}$ are evaluated with the random load, and $\hat{J}$ is obtained with the deterministic load, which is not computed based on $\hat{\mu}$ and $\hat{\sigma}$. In the robust designs with $\kappa = 1$ (Figures \ref{Fig_Ex1_Designs}b and \ref{Fig_Ex1_Designs}e), both methods produce similar designs with two split legs, and the design by AC-MDSA has slightly wider distances between the two legs and a slightly lower $\hat{J}(\bm{x}^*)$ (and lower $\hat{\mu}(\bm{x}^*)$ and $\hat{\sigma}(\bm{x}^*)$). In the robust designs with $\kappa=0.618$ (Figures \ref{Fig_Ex1_Designs}c and \ref{Fig_Ex1_Designs}f), both AC-MDSA and MC methods produce similar designs, and the design by AC-MDSA has a smaller distance between the two legs and a lower $\hat{J}(\bm{x}^*)$. This comparison verifies that, with only two samples in each optimization step, the entropic AC-MDSA produces similar designs and objective function values as the MC method with $1,000$ samples. We note that even though the MC achieves slightly higher objective function values, MC's solution can be potentially improved with more optimization steps and more computational time. Comparing designs with various $\kappa$ values, the design with higher weight in variance ($\kappa=0.618$) has wider legs and smaller $\hat{\sigma}(\bm{x}^*)$. In terms of computational efficiency, AC-MDSA generally has low computational costs as indicated in Table ~\ref{Tbl-Ex_1} due to its use of two samples.

\begin{figure}[H]
	\centering
	\includegraphics[width=5.0in]{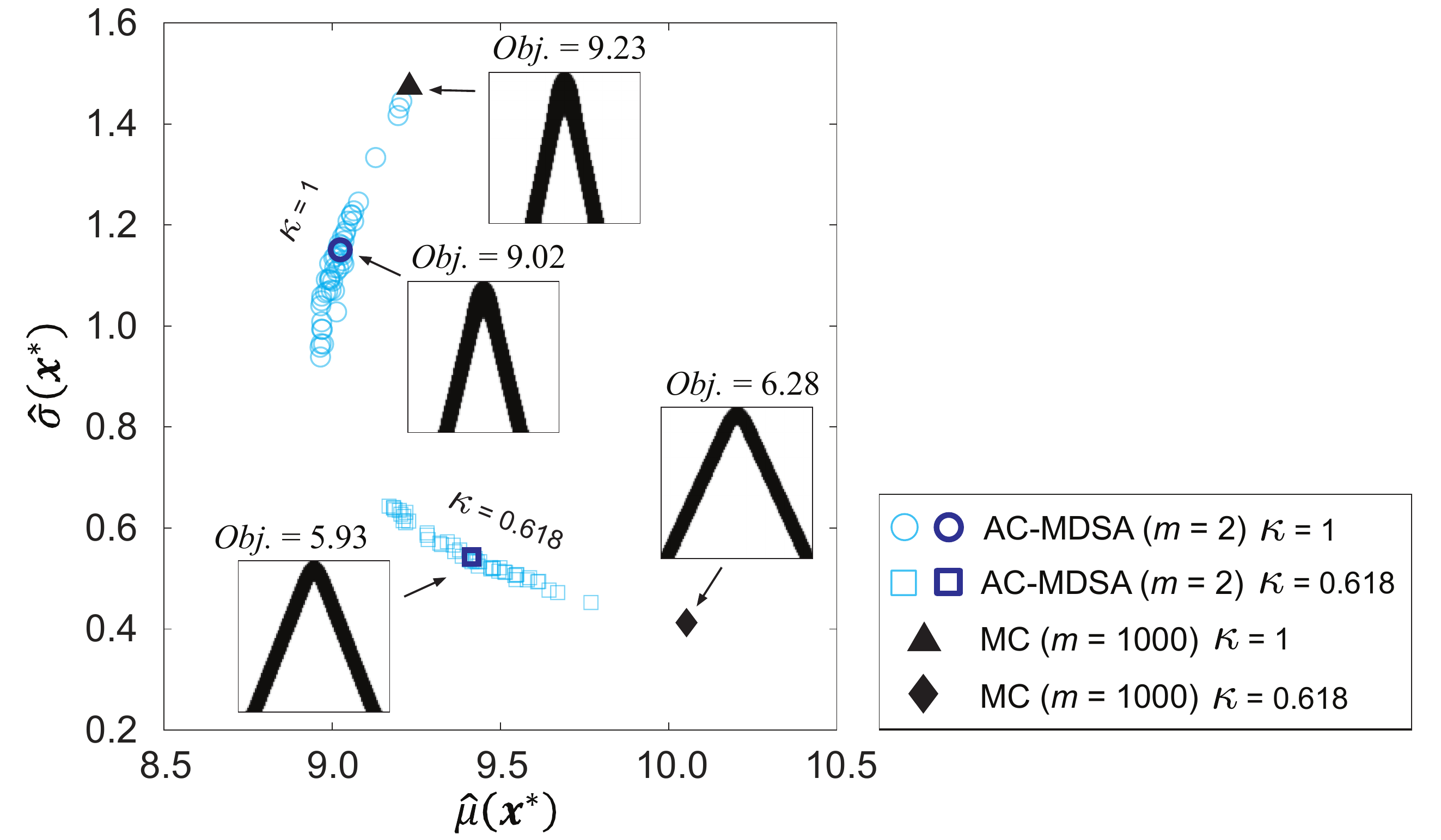}
	\caption{Performance comparison of the proposed AC-MDSA algorithm and the MC method: $\hat{\mu}(\bm{x}^*)$ versus $\hat{\sigma}(\bm{x}^*)$ for $\kappa = 1$ and $\kappa = 0.618$. AC-MDSA includes 50 trials for each $\kappa$ value. (Representative designs from each case is shown next to the highlighted markers.)}\label{Fig_Ex1_Stat}
\end{figure}
%
%\vspace*{-2.5in}
%\textcolor{red}{change to symbol for the figure}
\begin{figure}[H]
	\centering
	\includegraphics[width=5.5in]{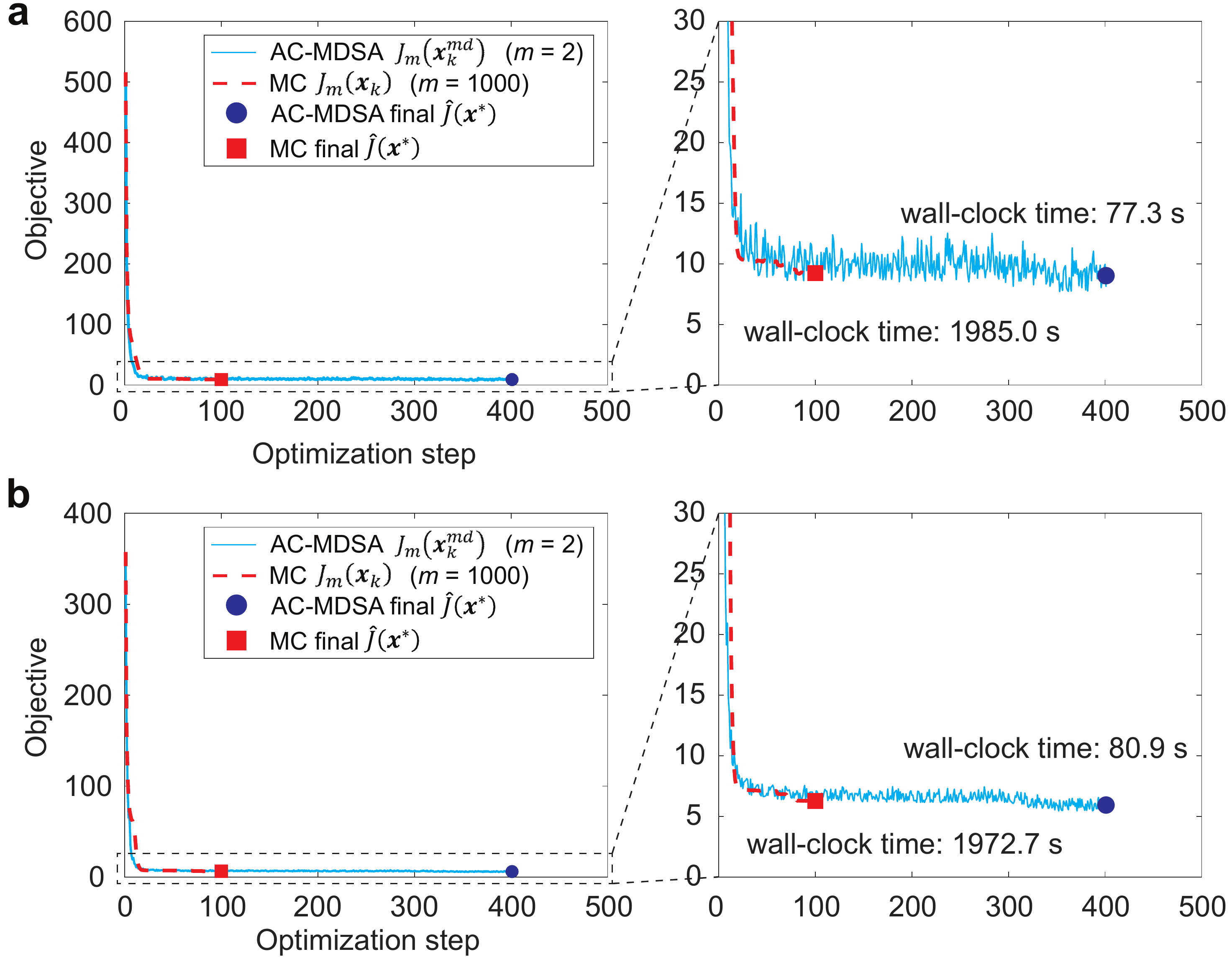}
	\caption{
		History of estimated objective function value for (a) $\kappa = 1$; (b) $\kappa = 0.618$. Highlighted markers represent $\hat{J}(\bm{x}^*)$.}\label{Fig_Ex1_Hist}
\end{figure}

To evaluate the overall performance and consistency of the entropic AC-MDSA, the $\hat{\mu}(\bm{x}^*)$ versus $\hat{\sigma}(\bm{x}^*)$ of the $50$ independent trials (one trial is one run of the numerical experiment) are plotted in Figure \ref{Fig_Ex1_Stat}. We observe that $50$ independent trials with $\kappa=1$ lead to similar $\hat{\mu}(\bm{x}^*)$, and those 50 trials with $\kappa = 0.618$ (higher weight on $\Var$) have similar $\hat{\sigma}(\bm{x}^*)$, indicating the AC-MDSA algorithm produces consistent designs. Also, the ones with $\kappa = 0.618$ have considerably lower $\hat{\sigma}(\bm{x}^*)$ and higher $\hat{\mu}(\bm{x}^*)$ than those with $\kappa = 1$, demonstrating the algorithm can effectively reflect the impact of $\kappa$ with only two samples. In the $\kappa = 0.618$ case, although MC method produces a design with the lowest $\hat{\sigma}(\bm{x}^*)$, its $\hat{\mu}(\bm{x}^*)$ is considerably higher than the designs produced by AC-MDSA, resulting in an overall higher objective function value.

Figure \ref{Fig_Ex1_Hist} shows the history of the estimated objective values of AC-MDSA and MC methods for $\kappa = 1$ and $\kappa = 0.618$. Note that the objective history of AC-MDSA is more oscillatory than the one of the MC method because the objective function in the entropic AC-MDSA is estimated with $m=2$ samples per step, and the one in the MC is estimated with $m=1000$ samples. However, the true objective of AC-MDSA evaluated at the end of the optimization with $10000$ samples has a similar value to that obtained by MC as indicated in Figure \ref{Fig_Ex1_Designs}.% Both AC-MDSA and MC methods experience a slow decrease in the estimated objective near the end, which is caused by the gradual reduction of filter radius.

%\vspace*{-1.5in}

\subsubsection{Study of sample size}
\begin{figure}[H]
	\centering
	\includegraphics[width=5.5in]{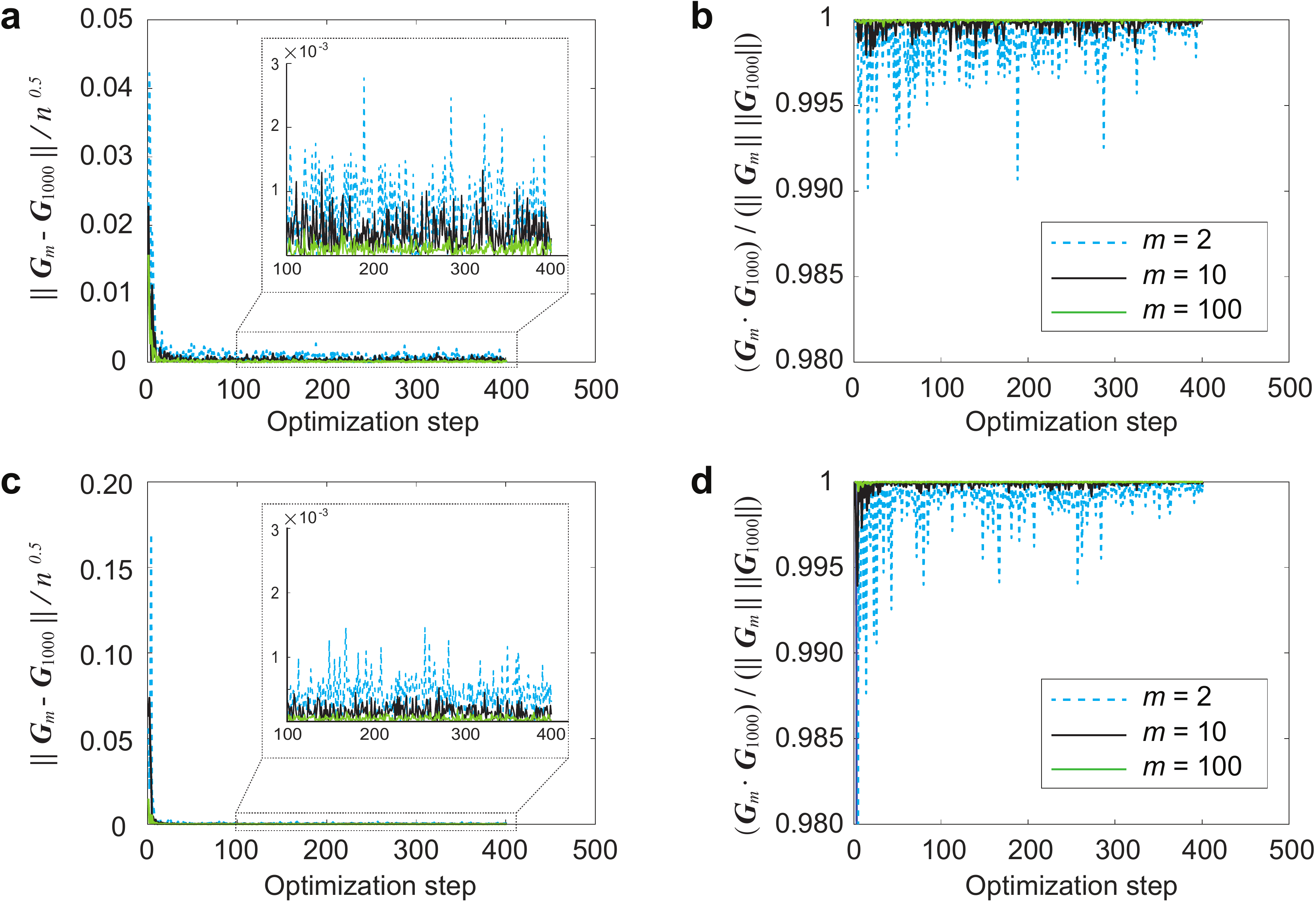}
	\caption{Error history comparison of stochastic gradients with $m=2, m=10, m=100$ samples used in AC-MDSA. (a) Error norm of the stochastic gradient: $\kappa = 1$; (b) cosine of the angle between the stochastic and the reference gradient vectors: $\kappa = 1$; (c) error norm of the stochastic gradient: $\kappa = 0.618$; (d) cosine of the angle between the stochastic and the reference gradient vectors: $\kappa = 0.618$.    }\label{Fig_Ex1_Error}
\end{figure}

%We now demonstrate the effectiveness of the step size recalibration scheme.
Next, we study the influence of various sample sizes $m$, which is used to compute the stochastic gradient, on the performance of the proposed AC-MDSA. We consider $m=2, m=10, m=100$ samples. 
%we study the accuracy level of the gradient estimator with various sample size $m$ and investigate the performance of the entropic AC-MDSA with respect to the different sample sizes. 
Figure \ref{Fig_Ex1_Error} shows the history of the error (norm) of stochastic gradients estimated using the three $m$ values for $\kappa = 1$ (Figures \ref{Fig_Ex1_Error} a and b) and $\kappa = 0.618$ (Figures \ref{Fig_Ex1_Error} c and d). The error is defined as the difference between the estimated gradient using $m$ samples and the reference estimated gradient using $1000$ samples. Several observations can be made. First, as we expect, a larger $m$ leads to a smaller difference between the estimated gradient and the reference estimated gradient. Second, the cosine of the angle between the stochastic and the reference estimated gradient vectors for both $\kappa = 1$ and $\kappa = 0.618$ are close to 1 after the first few steps, indicating the estimated gradient with a small sample size has fairly accurate directions, but this observation can be problem-dependent.
% as this example has a simple geometry that tends to have gradients less sensitive to the load randomness.

%Second, the error of the gradient estimator is larger in the cases of $\kappa = 0.618$ than the case of $\kappa = 1$ due to the larger weight of variance term in the objective function. Third, for both $\kappa$'s and all sample sizes, the magnitude of the error drops significantly after the first step size recalibration.

Various sample sizes $m$ produce gradient estimators with different accuracy levels; thus, we study the sensitivity of the AC-MDSA performance to $m$. Table ~\ref{Tbl-Ex_1} summarizes the performance and the associated computational cost of the entropic AC-MDSA with $m = 2$, $10$, and $100$ samples and compares with the ones from the MC method. The statistics in for the entropic AC-MDSA in Table ~\ref{Tbl-Ex_1} are averaged over the $50$ independent trials (for evaluating statistical consistency and are not needed in practice). The computational time shown in Table ~\ref{Tbl-Ex_1} is for reference. To have a more comprehensive comparison of computational cost, more investigation is needed.
%Table \ref{Tbl-Ex_1} demonstrates the high efficiency and effectiveness of the entropic AC-MDSA, and we have the following specific conclusions. 
For the entropic AC-MDSA, a larger $m$ leads to small differences in the final objective function values and convergence steps. This showcases that the proposed entropic AC-MDSA can perform high-quality updates with highly noisy gradient estimators (i.e., $m=2$). Therefore, we use $m=2$ for the remaining studies.
% of the remaining section as it is most efficient computationally. 

%% Table for ring example
%
%\textbf{Question: in the table, what is the Mean and s.t.d indicate? are they mu and sigma, or the averaged mu and averaged sigma, or mean and standard deviation for Obj.?}

\begin{table}[H]
	\centering
	\captionsetup{justification=centering}
	\caption{Performance of AC-MDSA (averaged over 50 trials) and MC methods: Simple column example}
	\label{Tbl-Ex_1}
	\resizebox{\textwidth}{!}{%
	\begin{tabular}{c c c c c c c c c c}
		\hline \hline
		Algorithm   & $\kappa$   	& {$\hat{J}(\bm{x}^*)$ } 	& {$\hat{\mu}(\bm{x}^*)$ } &  $\hat{\sigma}(\bm{x}^*)$   &  $N_\text{step}$ 	&  $N_\text{solve}$ & WC time 	& $\frac{\text{WC time}}{N_\text{step}}$ \\
		%					&        		&       	&            &  (avg.)   	&           		&  (avg.)         	&  (avg.)       & (avg.)   	& \\	
		&        		& (avg.) 	& (avg.)      & (avg.)     	&  (avg.)   		&  (avg.)        	& (sec.)         & (sec.)     \\ \hline
		AC-MDSA 	& 1 	  	& 9.02 	  	& 9.02          & 1.13         	&411.3         		& 862.6           		& 81.1 			& 0.2   		\\
		$m = 2$		 					& 0.618		& 5.93 	  	& 9.41          & 0.55         	& 420.9         		& 881.8          		& 82.8 			& 0.2   		\\

		\hline
		AC-MDSA 	& 1 	  	& 9.00 	  	& 9.00          & 1.07         	& 407.6         		& 855.3           		& 136.0 			& 0.3   		\\
		$m = 10$		 & 0.618		& 5.96 	  	& 9.48          & 0.52         	& 414.0         		& 868.1          		& 138.5 			& 0.3   		\\

		\hline
		AC-MDSA 	& 1 	  	& 8.99 	  	& 8.99          & 1.07         	& 403.4         		& 846.7           		& 807.3 			& 2.0   		\\
		$m = 100$	& 0.618		& 5.97 	  	& 9.49          & 0.52         	& 410.6         		& 861.2           		& 824.4				& 2.0   		\\

		\hline
		
		MC 		& 1 	  		& 9.23 	  	& 9.23          & 1.47         	& 100         		& $100000.0$           		& 1985.0 			& 19.9   		\\
		$m = 1000$	& 0.618 		& 6.28 		& 10.05          & 0.41       		& 100         		& $100000.0$      			& 1972.7 			& 19.3  		\\
		\hline
		
		\hline 	
		%\multicolumn{11}{l}{$^3$\footnotesize{}}
		%		Content\footnote{* Median value of 50 trials}
	\end{tabular}
}
\end{table}
%%%%%
%\noindent * Median value of 50 trials

%\noindent ** Values corresponding to the design with median objective function value

\subsubsection{Comparison of AC-MDSA and MDSA algorithms (with and without acceleration)}
\begin{figure}[H]
	\centering
	\includegraphics[width=6in]{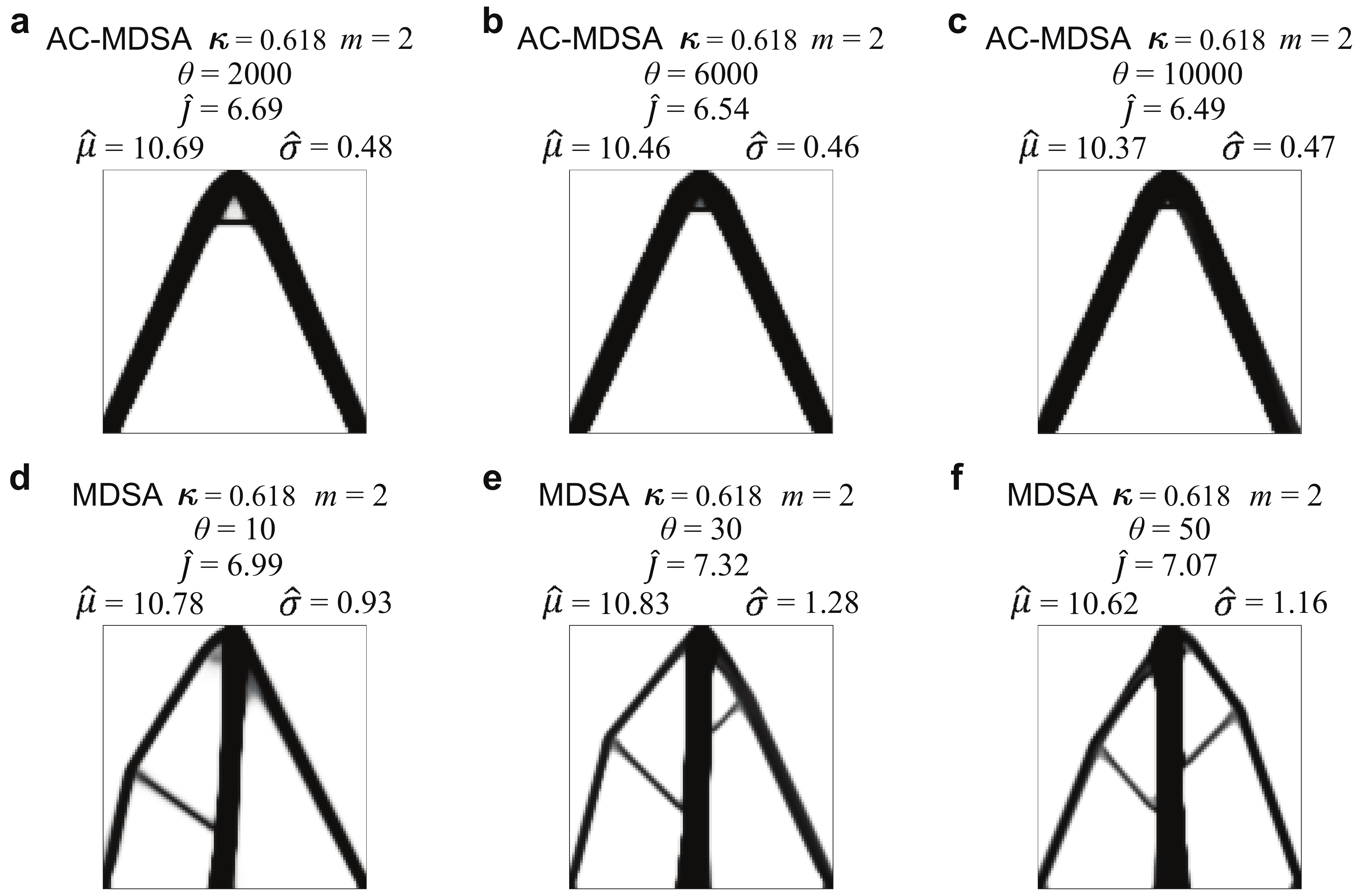}
	\caption{Final design of (a) AC-MDSA: $\theta = 2000$; (b) AC-MDSA: $\theta = 6000$;  (c) AC-MDSA: $\theta = 10,000$;  (d) MDSA: $\theta = 10$; (e) MDSA: $\theta = 30$; (f) MDSA: $\theta = 50$. The design in each case, respectively, is a representative design chosen from the 20 trials. }\label{Fig_ACSA_vs_MDSA}
\end{figure}

We compare the performance of the entropic AC-MDSA algorithm with the entropic MDSA algorithm (without acceleration) to demonstrate the advantage of the acceleration technique. In particular, we aim to demonstrate that, with the acceleration scheme, the AC-MDSA is less sensitive to various step sizes. We consider the case of $\kappa = 0.618$ and use the same step size recalibration, damping, and filter radius reduction setup for the MDSA algorithm. The symmetry of the designs is not imposed in this comparison. 
%The absence of the symmetry constraint requires smaller step size factors to improve the stability of the algorithm. 
For the entropic MDSA, the step size formula is adopted from \cite{Nemirovski2009,Zhang2020}. Figure \ref{Fig_ACSA_vs_MDSA} shows the final designs of AC-MDSA and MDSA with three values of step size scaling factor $\theta$. Notice $\theta$ is set to a smaller value than previous cases, and this is because when the symmetry constraint is absent, the algorithm needs a smaller step size to guarantee stable and steady convergence. Each design is a representative one selected from the results of $20$ independent trials. The range of $\theta$ value for the entropic MDSA is determined based on pilot runs. As shown in Figure \ref{Fig_ACSA_vs_MDSA}, the entropic MDSA is more sensitive to different choices of $\theta$ (i.e., different step sizes) than the entropic AC-MDSA. For various $\theta$ values considered, the entropic AC-MDSA yields similar results (which are also similar to Figures \ref{Fig_Ex1_Designs}c and f) with comparable performance, whereas the entropic MDSA yields less consistent results. Besides, although the design symmetry is not imposed, the entropic AC-MDSA produces nearly-symmetric designs while the entropic MDSA yields asymmetric ones, indicating the entropic AC-MDSA is more robust and stable than the entropic MDSA (without acceleration). Thus, the remaining of the study uses the entropic AC-MDSA algorithm.

%Figure \ref{Fig_ACSA_vs_MDSA} shows the final designs of AC-MDSA and MDSA with three step size factors. The design for each $\theta$ is selected from the ten consecutive trials and represents the majority of the 10 trials in geometric features. As can be seen, the AC-MDSA with three $\theta$ values produce designs . The AC-MDSA largely preserves the symmetry of the design even in the absence of symmetry constraint, which is a surprise. On the contrary, MDSA fail to converge to the correct designs because the designs have different topologies and gray members. Moreover, the designs of MDSA have higher objective function values than the AC-MDSA. The results also show that MDSA is less robust since designs with different $\theta$ are different, indicating high sensitivity to $\theta$. The comparison demonstrates the superior stability of the AC-MDSA over the conventional MDSA.

\subsection{Example 2: Half circle}
The second example demonstrates that the entropic AC-MDSA effectively captures the influence of various $\kappa$ values (relative weight of mean and variance for compliance) on the designs. 
%This example considers the half circle.
% subjected to a point load with a deterministic vertical component and random horizontal component. 
Figure \ref{Fig_Ex1_Ex2_geo}b shows the design domain and boundary conditions. The domain (discretized by $n=40,000$ polygonal elements \cite{Talischi2012}) is fixed on the outer perimeter and subjected to a point load that has a deterministic vertical component with magnitude 1 and random horizontal component $\sim \mathcal{N}(0,0.15^2) $. We consider three cases: $\kappa = 1$, $\kappa = 0.618$, and $\kappa = 0.282$. The filter radius $R$ is initialized as $0.03$ and reduced to $0.004$ after $300$ steps with an interval of 30 steps. We choose $\theta = 8000n$, $\theta = 100n$, and $\theta = 10n$ for $\kappa = 1$, $\kappa = 0.618$, and $\kappa = 0.282$, respectively; $N_{rst} = 300$, $N_{damp} = 450$,  $\epsilon_{damp}=0.05$, $N_{max} = 600$, and $N_{min}=450$.

\begin{figure}[H]
	\centering
	\includegraphics[width=6.8in]{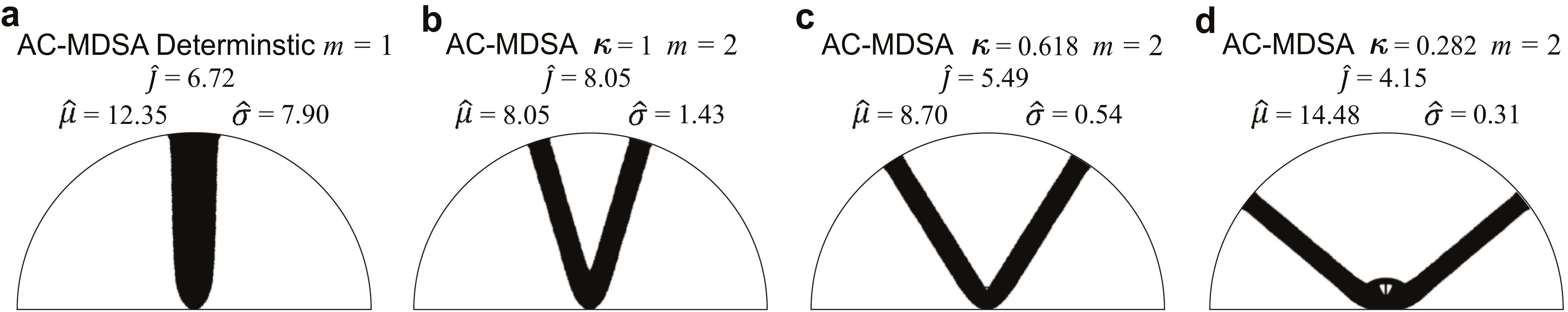}
	\caption{Final designs and objective function values of half circle: (a) deterministic;  (b) $\kappa = 1$; (c) $\kappa = 0.618$; (d) $\kappa = 0.282$. The design in (b), (c), and (d), respectively, is a representative design chosen from the 50 trials.}\label{Fig_Ex2_Designs}
\end{figure}

Figure \ref{Fig_Ex2_Designs} shows the designs obtained by the entropic AC-MDSA for the deterministic case (i.e., the horizontal load is $0$) and three stochastic cases with a wide range of $\kappa$. For the stochastic cases, each design is a representative one from $50$ independent trials with the objective function values close to the mean of the 50 objective function values. The three stochastic designs show the impact of various $\kappa$ values: as $\kappa$ decreases (more weight on the variance), the angle between the two arms increases, improving the robustness in resisting the random horizontal load.

\begin{figure}[H]
	\centering
	\includegraphics[width=5in]{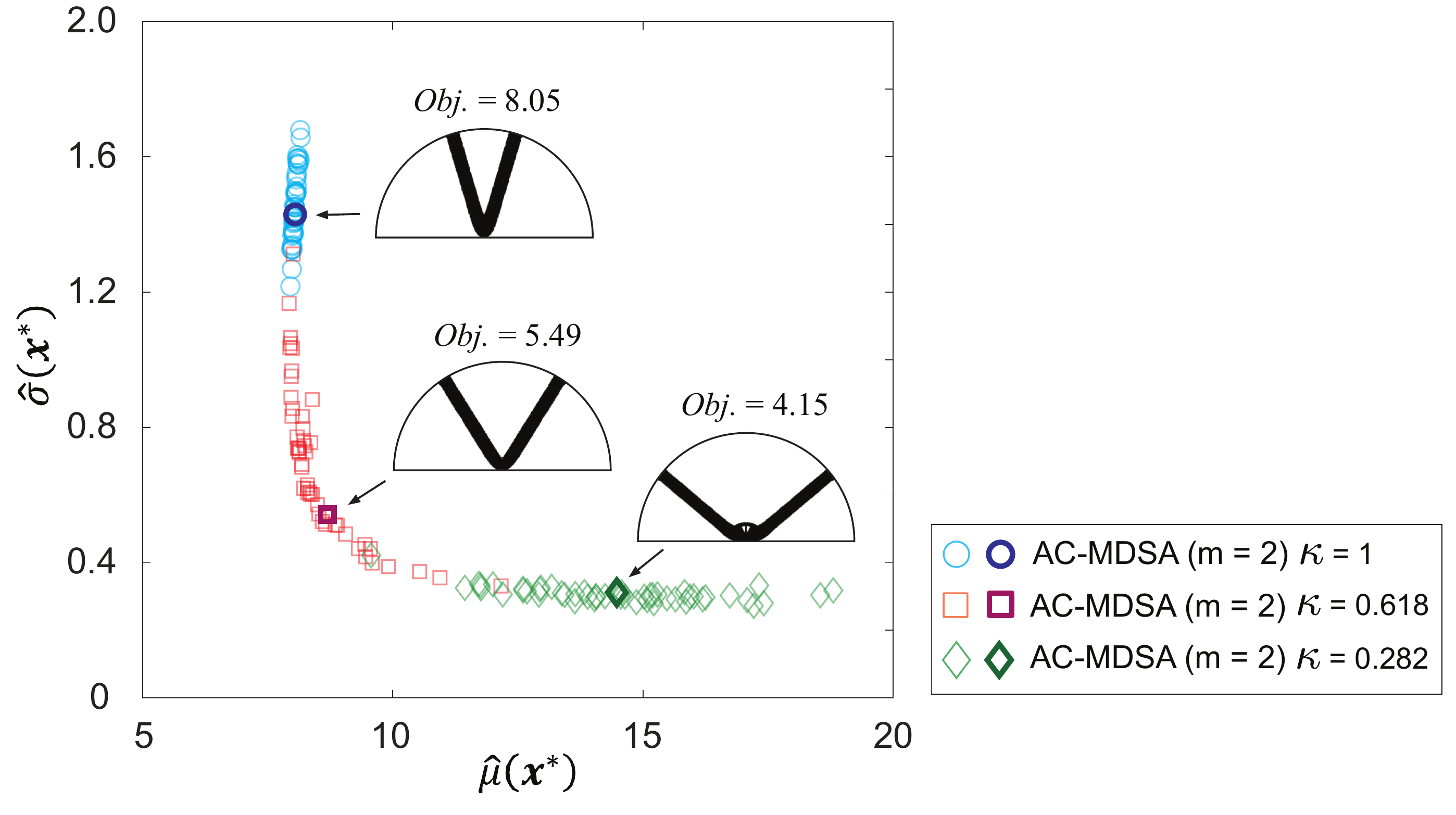}
	\caption{$\hat{\mu}(\bm{x}^*)$ versus $\hat{\sigma}(\bm{x}^*)$ of the 50 trials by the entropic AC-MDSA. (Highlighted markers correspond to the presented designs.)}\label{Fig_Ex2_Stat}
\end{figure}

%The values of $\hat{\mu}(\bm{x}^*)$ versus $\hat{\sigma}(\bm{x}^*)$ reflect the impact of varying $\kappa$, as shown in Figure \ref{Fig_Ex2_Stat}. As expected, as $\kappa$ decreases, $\hat{\mu}(\bm{x}^*)$  increases and $\hat{\sigma}(\bm{x}^*)$ decreases. 
The impact of varying $\kappa$ is shown in Figure \ref{Fig_Ex2_Stat}, which plots $\hat{\mu}(\bm{x}^*)$ versus $\hat{\sigma}(\bm{x}^*)$ of a total of $150$ independent trials ($50$ for each $\kappa$) with representative designs. Several observations can be made. First, as $\kappa$ decreases, $\hat{\sigma}(\bm{x}^*)$ decreases (indicating improved robustness) and $\hat{\mu}(\bm{x}^*)$  increases. The AC-MDSA produces consistent designs for each $\kappa$ case. Second, the designs for larger $\kappa$ typically have similar $\hat{\mu}(\bm{x}^*)$ but widely distributed $\hat{\sigma}(\bm{x}^*)$, while the designs for smaller $\kappa$ typically have similar $\hat{\sigma}(\bm{x}^*)$ but widely distributed $\hat{\mu}(\bm{x}^*)$. This observation is consistent with the definition of the objective function in \eqref{estimated_obj}.

%a smaller variation in the component in the results with higher weight in objective.This is consistent with the relative importance of mean and variance as defined in the objective function. 

\subsection{Example 3: Robust designs with multiple random loads}

\begin{figure}[H]
	\centering
	\includegraphics[width=5in]{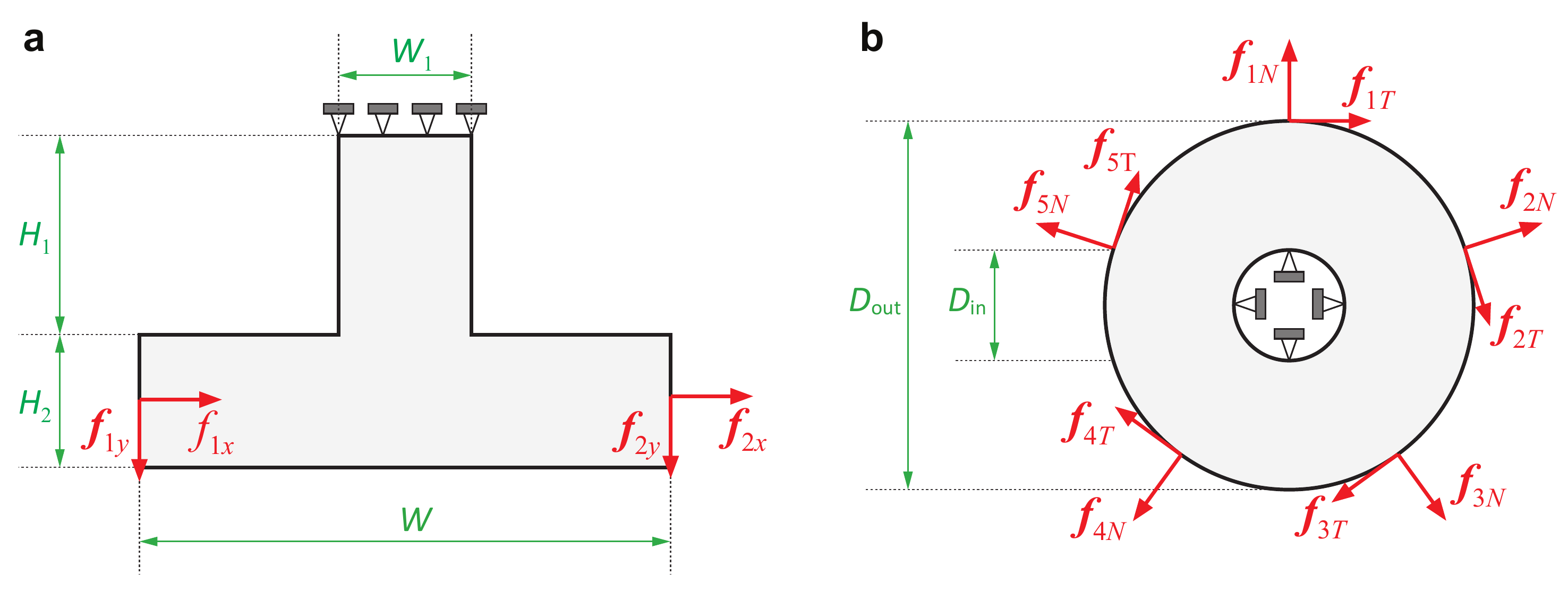}
	\caption{Geometry and boundary conditions of (a) double hook, $W = 4, W_1 = 1, H_1 = 1, H_2 = 1.5$, two point loads have deterministic vertical components $\boldsymbol{f}_{1y} = \boldsymbol{f}_{2y} = 1$ and random horizontal components $\boldsymbol{f}_{1x},\boldsymbol{f}_{2x} \sim \mathcal{N}(0,0.1^2)$; (b) disk, $D_{out} = 2, D_{in} = 0.6$, five point loads have deterministic normal components $\boldsymbol{f}_{1N}=\boldsymbol{f}_{2N}=\boldsymbol{f}_{3N}=\boldsymbol{f}_{4N}=\boldsymbol{f}_{5N}=1$ and random tangential components $\boldsymbol{f}_{1T},\boldsymbol{f}_{2T},\boldsymbol{f}_{3T},\boldsymbol{f}_{4T},\boldsymbol{f}_{5T}\sim \mathcal{N}(0,0.1^2)$.}\label{Fig_Ex3_Ex4_geo}
\end{figure}

The third example, which includes the double hook and the disk problem, is designed to show that the AC-MDSA algorithm can tackle problems with various problem sizes, geometries, and multiple independent random loads. Additionally, using the double hook example, we show that the parameters of the AC-MDSA algorithm are insensitive to various mesh sizes.
Figure \ref{Fig_Ex3_Ex4_geo} shows the design domains and boundary conditions of the double hook and the disk problems. 
%The dimensions of the double hook are $W = 4$, $W_1 = 1$, $H_1 = 1$, $H_2 = 1.5$. 

\begin{figure}[H]
	\centering
	\includegraphics[width=6in]{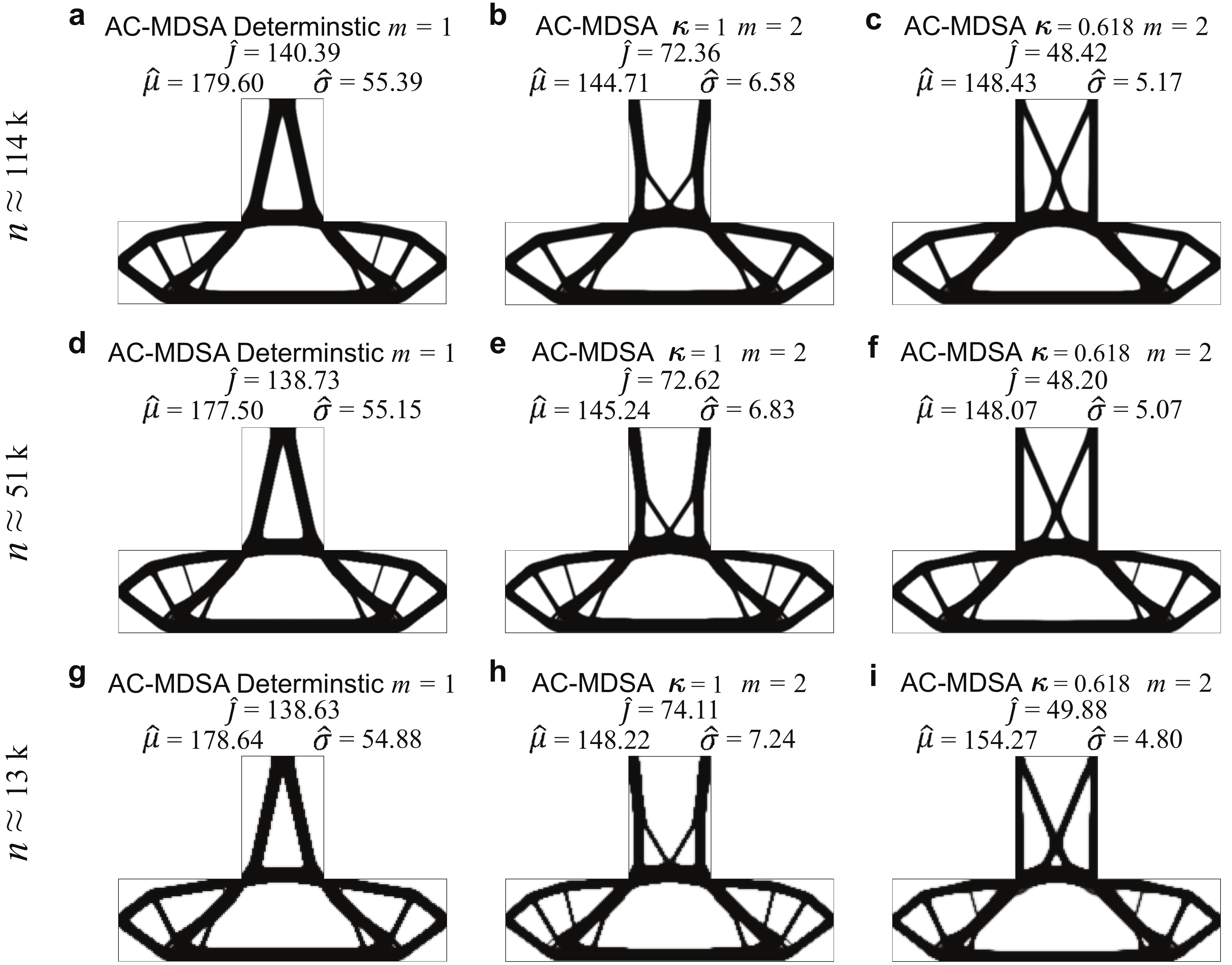}
	\caption{Double hook: deterministic and robust designs ($\kappa = 1$ and $\kappa=0.618$) obtained by the AC-MDSA algorithm: (a)-(c) $n = 114,048$, (d)-(f) $n = 50,688$, (g)-(i) $n = 12,672$. The design of (b), (c), (e), (f), (h), and (i), respectively, is a representative design chosen from the 50 trials.}\label{Fig_Ex3_Designs}
\end{figure}
\subsubsection{Double hook}
In the double hook problem, 
%two point loads are applied at the mid-height at both ends. 
the two point loads have deterministic vertical components with magnitudes 1 and random horizontal components $\sim \mathcal{N}(0,0.1^2)$. 
%The domain of the disk problem has an outer diameter $D_{out} = 2$ and inner diameter $D_{in} = 0.6$. 
%
We use $\theta = n$ and $\theta = 0.03n$ for $\kappa = 1$ and $\kappa=0.618$, respectively, and $N_{rst} = 100$, $N_{damp} = 450$, $\epsilon_{damp} = 0.075$, $N_{max} = 600$, and $N_{min} = 450$.
We first evaluate the sensitivity of the AC-MDSA algorithmic parameters (e.g., step size factor $\theta$ and initial step to monitor recalibration $N_{rst}$) to various mesh sizes, $n = 114,048$, $n = 50,688$ and $n = 12,672$.
% with the same algorithmic parameters (same rule to compute $\theta$ in terms of $n$). 
For comparison, we also solve the problem using MC method with 1000 samples.

\begin{figure}[H]
	\centering
	\includegraphics[width=6in]{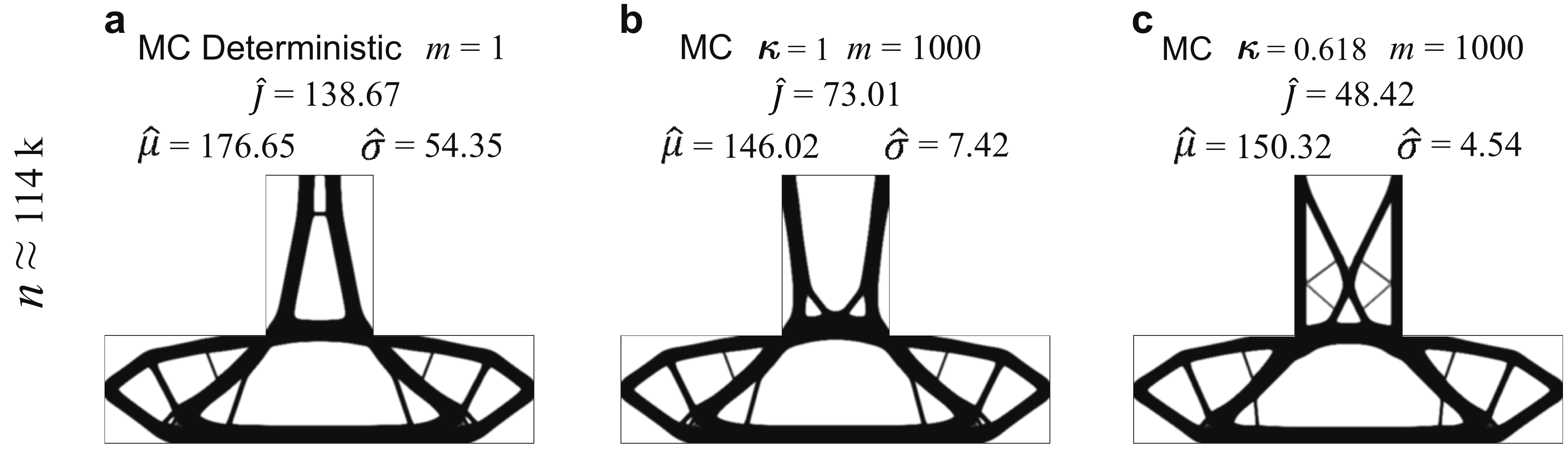}
	\caption{Double hook ($n = 114,048$): deterministic and robust designs ($\kappa = 1$ and $\kappa=0.618$) obtained by MC method with 1000 samples.}\label{Fig_Ex3_MCMMA_Designs}
\end{figure}
Comparing (vertically) the designs with three problem sizes, as shown in Figure \ref{Fig_Ex3_Designs}, they have consistent geometric features and similar objective function values for each case of $\kappa$. This observation demonstrates that the proposed AC-MDSA algorithm and associated parameters can lead to mesh-insensitive designs. 
Consistent observations can be made among designs from various problem sizes. The deterministic and robust designs differ in the upper domain. The deterministic design forms a single connection to the support, resulting in less resistance to moments and horizontal loads. The robust design with $\kappa = 1$ has two separated arms without braces, which can carry moment but is weak in resisting horizontal shear forces. The robust design with $\kappa = 0.618$ forms a brace with separated arms, indicating an improved strength to resist the stochastic lateral load. The increase in robustness is also revealed in the decrease in $\hat{\sigma}(\bm{x}^*)$ of the three designs from left to right.

\begin{figure}[H]
	\centering
	\includegraphics[width=6in]{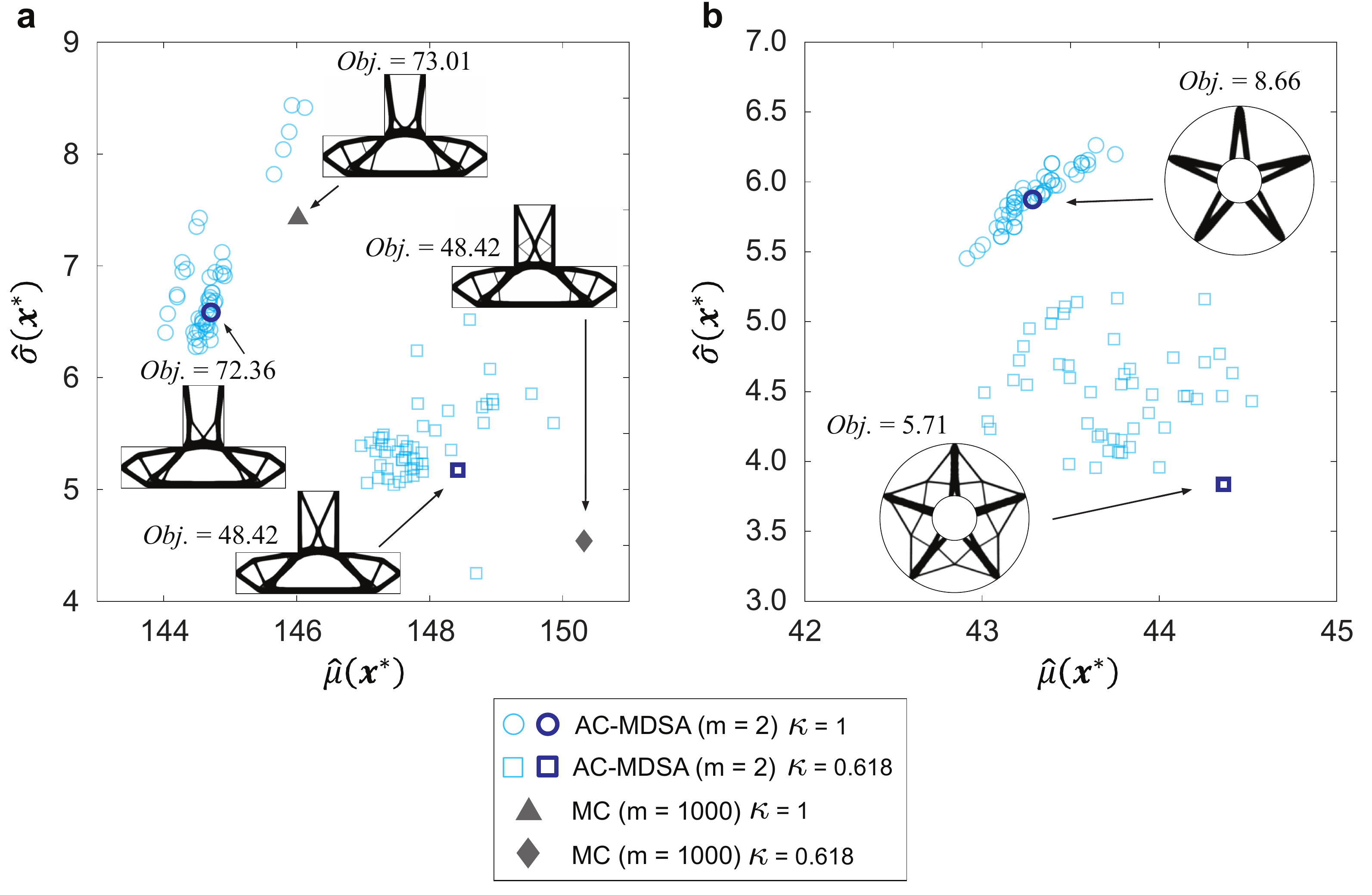}
	\caption{$\hat{\mu}(\bm{x}^*)$ versus $\hat{\sigma}(\bm{x}^*)$ of the 50 trials by AC-MDSA (highlighted markers correspond to the presented designs) (a) double hook (including the two designs by MC method); (b) disk.}\label{Fig_Ex4_Stat}
\end{figure}

The final designs and objective function values obtained by the MC method with 1000 samples are shown in Figure \ref{Fig_Ex3_MCMMA_Designs}. The main geometric features are similar to the designs from the AC-MDSA with two samples, but with more small branches. For the objective function values, AC-MDSA achieves a slightly lower value in the $\kappa = 1$ design and an identical value in the $\kappa = 0.618$ design as compared to the MC method. 
%The comparison with MC method justifies the effectiveness of the two-sample AC-MDSA to reflect the impact of $\kappa$ and producing robust designs.
Figure \ref{Fig_Ex4_Stat} (a) shows $\hat{\mu}(\bm{x}^*)$ and $\hat{\sigma}(\bm{x}^*)$ of the 50 trials from the AC-MDSA and one trial from the MC method. 
%, the data points associated with each $\kappa$ value are , and the makers representing the same $\kappa$ are grouped in the same areas. This observation again shows the statistical consistency of AS-MDSA. 
We can observe that a lower $\kappa$ value produces designs with lower $\hat{\sigma}(\bm{x}^*)$.
The statistics, including computational cost, is shown in Table ~\ref{Tbl-Ex_3}, and the data related to AC-MDSA are averaged values over the 50 trials (for evaluating statistical consistency and are not needed in practice). The AC-MDSA algorithm solves approximately 1240 linear systems with an average wall-clock time of approximately 2.4 seconds per step. %The total wall-clock time of the MC method is more than 15 times longer than AC-MDSA, even though the MC method uses 1/6 of the total steps of AC-MDSA. %This comparison shows that the AC-MDSA algorithm, which requires two linear system solves per step, reduces the computational cost of RTO problems drastically.

%% Table for double hook example
\begin{table}[H]
	\centering
	\captionsetup{justification=centering}
	\caption{Performance of AC-MDSA (averaged over 50 trials) and MC methods: double hook example ($n = 114,048$) }
	\label{Tbl-Ex_3}
		\resizebox{\textwidth}{!}{%
	\begin{tabular}{c c c c c c c c c}
		\hline \hline 
		Algorithm   & $\kappa$   	& {$\hat{J}(\bm{x}^*)$ } 	& {$\hat{\mu}(\bm{x}^*)$ } &  $\hat{\sigma}(\bm{x}^*)$  &  $N_\text{step}$ 	&  $N_\text{solve}$ & WC time 	& $\frac{\text{WC time}}{N_\text{step}}$ \\
		%					&        		&       	&            &  (avg.)   	&           		&  (avg.)         	&  (avg.)       & (avg.)   	& \\	
		&        		& (avg.)      	& (avg.)          & (avg.)          	& (avg.)          		& (avg.)                	& (sec.)       & (sec.)    \\ \hline
		AC-MDSA 	& 1 	  		& 72.36 	  	& 144.72          & 6.82         	& 600.0         		& 1240.0           		& 1539.2 			& 2.6    		\\
		$m = 2$ & 0.618		& 48.50 	  	& 147.88          & 5.41         	& 600.0         		& 1240.0           		& 1291.2			& 2.2   		\\
		\hline
		MC  		& 1 	  		& 73.01 	  	& 146.02          & 7.42         	& 100         		& $10^5$           		& 24129 			& 2412.9   		\\
		$ m = 1000$ & 0.618 		& 48.42 		& 150.32          & 4.54       		& 100         		& $10^5$      			& 24315 			& 2431.5  		\\
		\hline
		
		\hline 	
		%\multicolumn{11}{l}{$^3$\footnotesize{}}
	\end{tabular}
}
\end{table}
%%%%%
\subsubsection{Disk}
\begin{figure}[H]
	\centering
	\includegraphics[width=6in]{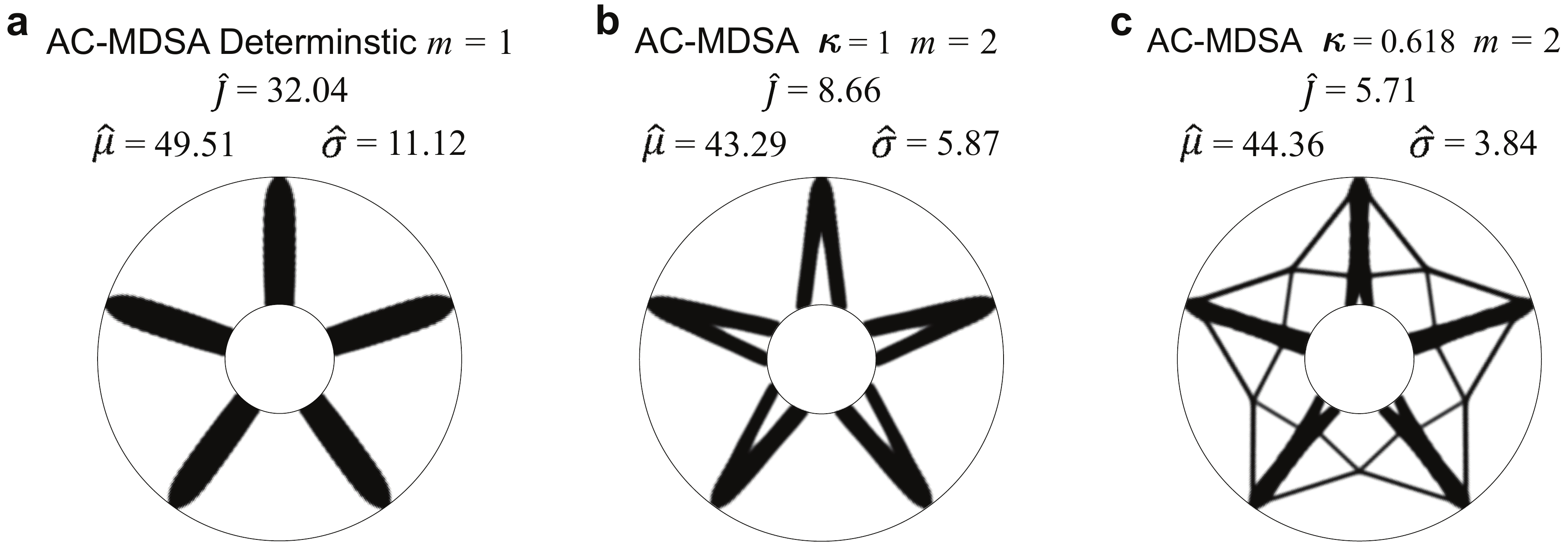}
	\caption{Final designs and objective function values of disk by AC-MDSA: (a) deterministic;  (b) $\kappa = 1$; (c)$\kappa = 0.618$. The design in (b) and (c), respectively, is a representative design chosen from the 50 trials.}\label{Fig_Ex4_Designs}
\end{figure}

In the disk problem, five loads are equally distributed on the outer perimeter, and each has a deterministic normal component 1 and a random tangential component $\sim \mathcal{N}(0,0.1^2)$. We use $n=72,000$ elements, $\theta = 3000n$ and $\theta = 0.1n$ for $\kappa = 1$ and $\kappa=0.618$, respectively, $N_{rst} = 300$, $N_{damp} = 450$,  $\epsilon_{damp} = 0.05$, $N_{max} = 600$, and $N_{min} = 450$.
Figure \ref{Fig_Ex4_Designs} shows the representative optimized designs for deterministic, $\kappa = 1$, and $\kappa = 0.618$ cases. The deterministic design contains rods with uniform widths, resisting normal load components, whereas the robust design with $\kappa = 1$ leads to a structure with two branches that resist each random tangential loads. In the robust design with $\kappa = 0.618$, lateral braces are formed to further enhance the resistance of random tangential loads. The $\hat{\sigma}(\bm{x}^*)$ values of the three designs confirm that the robustness is effectively improved when $\kappa$ drops. Table ~\ref{Tbl-Ex_4} and Figure \ref{Fig_Ex4_Stat} (b) show the statistics of the 50 trials from AC-MDSA, which solves two linear systems per step with an average wall-clock time of 1.7 seconds. 

% shows the $\hat{\mu}(\bm{x}^*)$ and $\hat{\sigma}(\bm{x}^*)$ of the 50 trials for the double hook and disk problem by AC-MDSA, respectively. The result by MC method for the double hook is also included.  We can observe that, for both problems, the markers representing different $\kappa$ values are widely separated, and the makers representing the same $\kappa$ are grouped in the same areas. This observation again proves the statistical consistency of AS-MDSA. Also, a lower $\kappa$ produces designs with lower $\hat{\sigma}(\bm{x}^*)$.

%% Table for disk example
\begin{table}[H]
	\centering
	\captionsetup{justification=centering}
	\caption{Performance of AC-MDSA (averaged over 50 trials): Disk example }
	\label{Tbl-Ex_4}
		\resizebox{\textwidth}{!}{%
	\begin{tabular}{c c c c c c c c c}
		\hline \hline
		Algorithm   & $\kappa$   	& {$\hat{J}(\bm{x}^*)$ } 	& {$\hat{\mu}(\bm{x}^*)$ } &  $\hat{\sigma}(\bm{x}^*)$ &  $N_\text{step}$ 	&  $N_\text{solve}$ & WC time 	& $\frac{\text{WC time}}{N_\text{step}}$ \\
		%					&        		&       	&            &  (avg.)   	&           		&  (avg.)         	&  (avg.)       & (avg.)   	& \\	
		&        		&(avg.)     	&(avg.)            &(avg.)           	&(avg.)           		&(avg.)                 	& (sec.)         & (sec.)     \\ \hline
		AC-MDSA 	& 1 	  		& 8.66 	  	& 43.29          & 5.89         	& 600.0         		& 1240.0           		& 1047.8 			& 1.7   		\\
		$m = 2$ & 0.618		& 5.72 	  	& 43.74          & 4.51         	& 600.0         		& 1240.0           		& 1030.6 			& 1.7   		\\
		%		\hline
		%		\multirow{2}{*}{MC -1000} 		& 1 	  		& 9.71 	  	& 9.71          & 2.30         	& 100         		& $10^5$           		& 8094 			& 80.94   		\\
		%		& 0.618 		& 6.59 		& 10.42          & 0.64       		& 100         		& $10^5$      			& 8228 			& 82.28  		\\
		\hline
		
		\hline 	
		%\multicolumn{11}{l}{$^3$\footnotesize{}}
	\end{tabular}
}
\end{table}
%%%%%
%\addtocounter{footnote}{-1}
%\footnotetext{Median value}
%\addtocounter{footnote}{1}
%\footnotetext{Values corresponding to median value}
%\noindent * Median value of 50 trials
%
%\noindent ** Values corresponding to the design with median objective function value

Both the double hook and disk examples show that the two-sample AC-MDSA can solve problems with various mesh sizes, complex geometries, and multiple independent random loads. As $\kappa$ changes, we observe apparent changes in both the designs and $\hat{\sigma}(\bm{x}^*)$ values. The optimized designs and algorithmic parameters are insensitive to the change of mesh sizes. Finally, we show that the AC-MDSA algorithm requires a low computational cost to handle RTO problems as it needs only two linear solves per step.

\subsection{Example 4: three-dimensional crane}
\begin{figure}[H]
	\centering
	\includegraphics[width=2.5in]{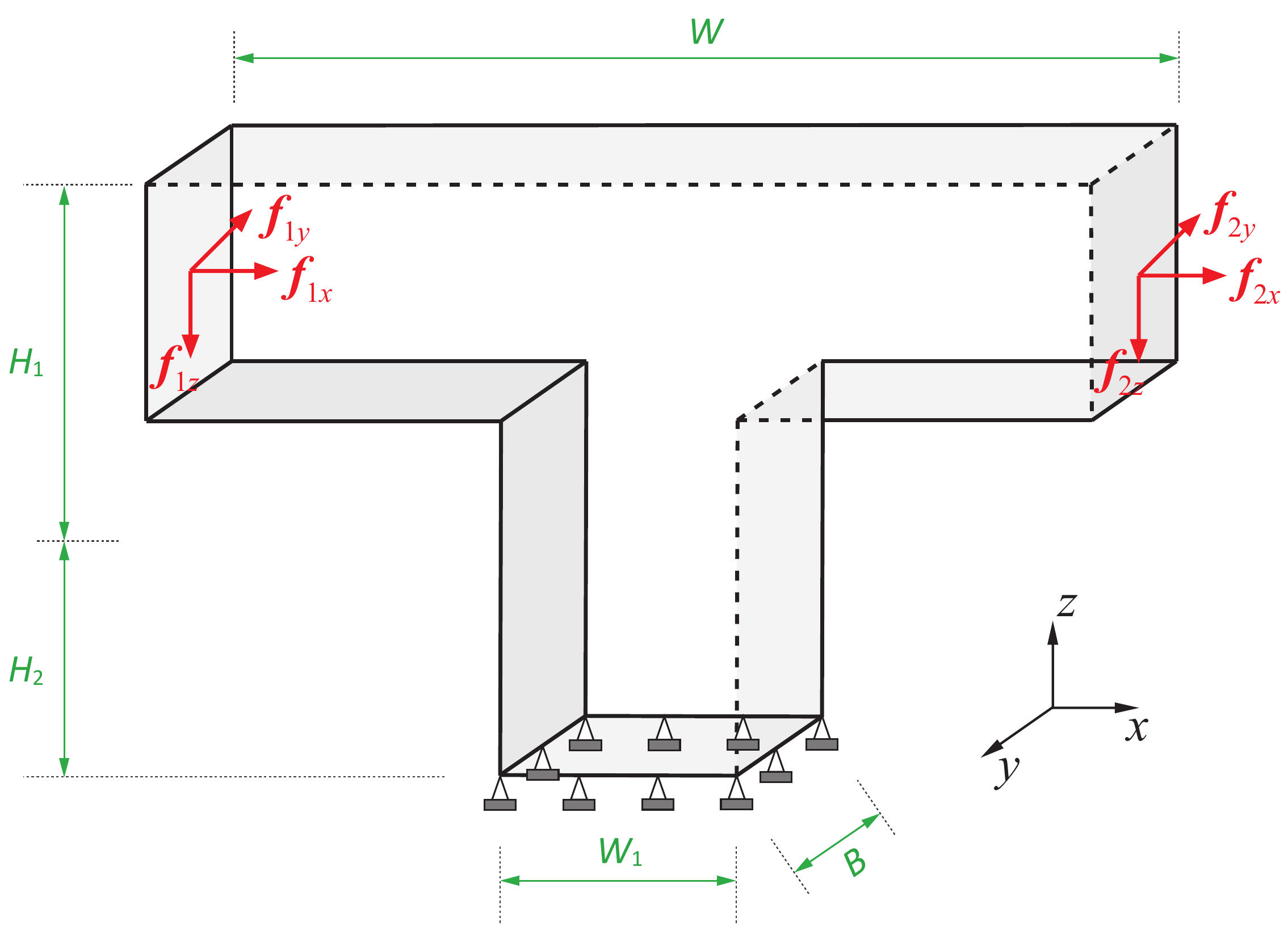}
	\caption{Geometry and boundary conditions of Example 4: 3D crane, $W = 4, W_1 = 1, H_1 = 1.5, H_2 = 1, B = 1$, two point loads have deterministic $z$ components $\boldsymbol{f}_{1z} = \boldsymbol{f}_{2z} = 1$ and random $x$ and $y$ components $\boldsymbol{f}_{1x},\boldsymbol{f}_{1y},\boldsymbol{f}_{2x},\boldsymbol{f}_{2y} \sim  \mathcal{N}(0,0.1^2). $} \label{Fig_Ex5_geo}
\end{figure}

The last example, which solves a 3D crane problem, demonstrates the applicability and efficiency of the entropic AC-MDSA. Figure \ref{Fig_Ex5_geo} shows the domain and boundary conditions.
% The dimensions are $W = 4$, $W_1 = 1$, $H_1 = 1.5$, $H_2 = 1$, and $B = 1$. 
The domain is fixed on the top and subjected to two point loads that have deterministic $z$ components with magnitudes 1 and random $x$ and $y$ components $\sim \mathcal{N}(0,0.1^2) $. The FE mesh consists of $n = 352,000$ hexahedral elements. The filter radius $R$ is initialized as $0.15$ and starts to decrease after $320$ steps by $1/30$ every 25 steps until $0.05$. We use $\theta = 2000$ and impose symmetry constraints with respect to the $x$ and $y$ planes, and $N_{rst} = 100$, $N_{damp} = 430$, $\epsilon_{damp} = 0.05$, $N_{max} = 450$, and $N_{min} = 430$. The objective function values of the final designs are evaluated using $1,000$ samples.
%This 3D problem requires iterative linear solvers. 
We use the GPU-accelerated preconditioned conjugate gradient (PCG) built-in solver from Matlab with the Jacobi preconditioner and choose a relatively high tolerance of $10^{-4}$ for convergence as the entropic AC-MDSA does not require accurate evaluation of sensitivity.
%\vspace*{-1in}
%
%
%\vspace*{-1.5in}
\begin{figure}[H]
	\centering
	\includegraphics[width=5in]{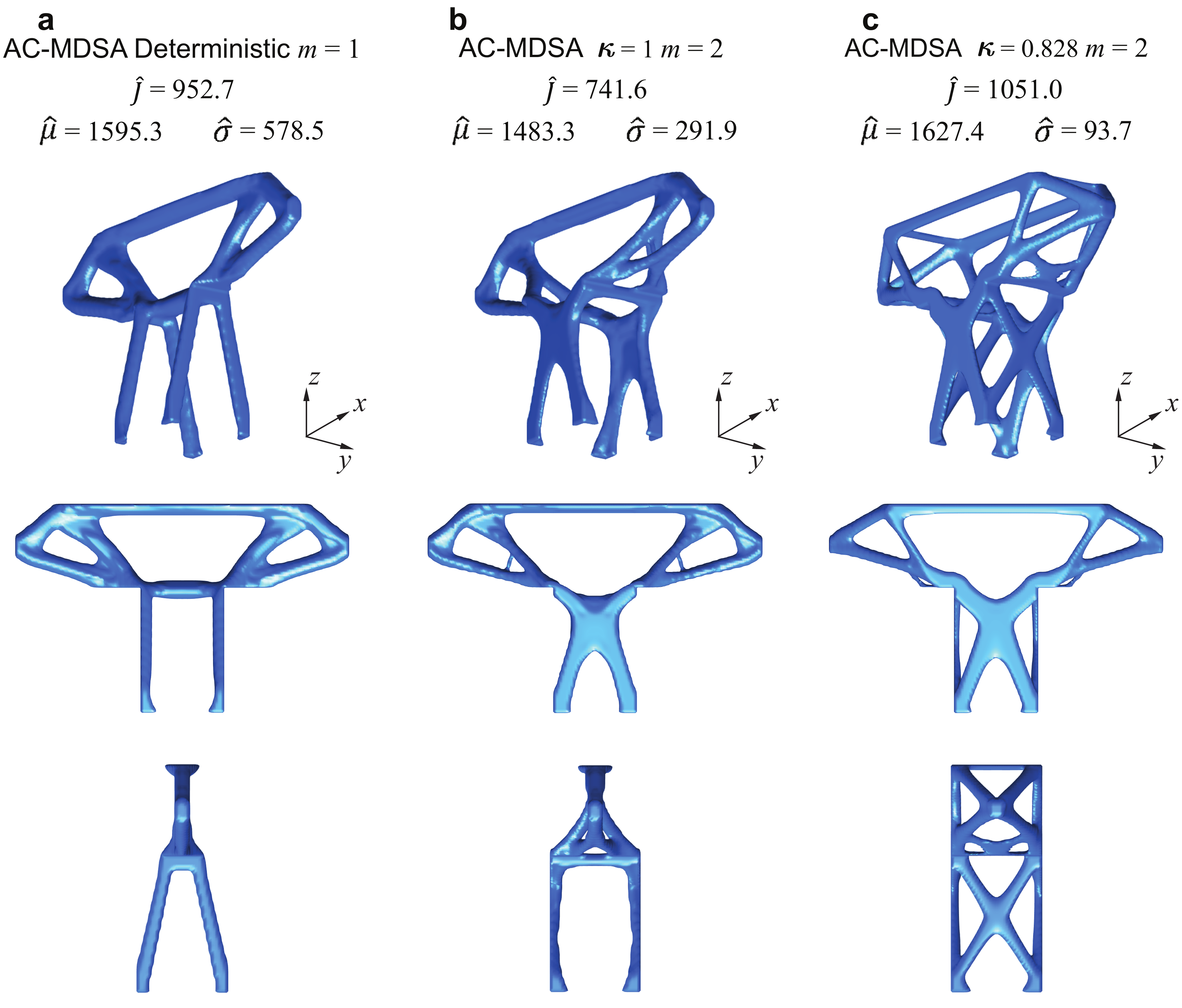}
	\caption{Optimized designs and objective function values of 3D crane from AC-MDSA: (a) deterministic;  (b) $\kappa = 1$; (c) $\kappa = 0.828$. The design in (b) and (c), respectively, is a representative design chosen from the 10 trials. }\label{Fig_Ex5_Designs}
\end{figure}

The optimized designs and objective function values for deterministic, $\kappa = 1$, and $\kappa = 0.828$ cases are shown in Figure \ref{Fig_Ex5_Designs}. We observe that the entropic AC-MDSA algorithm captures the influence of different $\kappa$ values on the final designs, both qualitatively and quantitatively. Qualitatively, in the deterministic design (Figure \ref{Fig_Ex5_Designs}a), no lateral braces are formed among the four columns on the upper part, resulting in poor resistance to shear in the $x$-direction and torque in the $x - y$ plane. In addition, the material in the lower part is mostly distributed within the $x - z$ plane, which also leads to poor resistance to loads in the $y$-direction. The robust design with $\kappa = 1$ (Figure \ref{Fig_Ex5_Designs}b), on the other hand, forms pairs of braces in the $x - z$ planes between the four columns, which improves the resistance to the random load components in the $x$- and $y$-directions that potentially impose shear and torsion. However, no braces appear in the $y - z$ planes. In the lower part, two branches are formed in the upper and middle chords of the beam. These branches can increase the stiffness of resisting the random loads in the $y$-direction. Finally, the robust design with $\kappa = 0.828$ (Figure \ref{Fig_Ex5_Designs}c) forms four braces in both the $x - z$ and $y - z$ planes between the four columns, leading to the highest resistance to the shear and torsion imposed by the random load components in the $x-$ and $y$-directions. In the lower part of the design, the upper chord branches are further split to enhance the resistance to loads in the $y$-direction. The middle chord becomes two independent members, and the lower chord splits into two branches. In addition, two members connecting the two lower chords are formed. These features clearly indicate the increase in the structural robustness when $\kappa$ decreases. Quantitatively, the influence of $\kappa$ is also revealed by the values of $\hat{\mu}(\bm{x}^*)$ and $\hat{\sigma}(\bm{x}^*)$ of the optimized designs. For the deterministic case, the design has both the highest $\hat{\mu}(\bm{x}^*)$ and $\hat{\sigma}(\bm{x}^*)$ because the load randomness is not considered in the optimization. For the robust designs, as $\kappa$ becomes smaller, $\hat{\mu}(\bm{x}^*)$ increases while $\hat{\sigma}(\bm{x}^*)$ decreases considerably, which is consistent with the corresponding importance in the objective function of the RTO formulation \eqref{eq:RTO original form.}.

This 3D example shows that the proposed AC-MDSA algorithm effectively produces designs with various levels of robustness. The AC-MDSA uses a relatively high tolerance for the iterative linear solve, which may suggest high tolerance can be used to reduce computational cost further as AC-MDSA does not require accurate evaluation of gradients. However, more investigation is needed to verify this potential.

%%%%%%%%%%%%%%%%%%%%%%%%%%%%%%%%%%%%%%%%%% CONCLUSION %%%%%%%%%%%%%%%%%%%%%%%%%%%%%%%%%%%%%%%%%%%%%%%%%%
\section{Concluding remarks}

%Load randomness is one of the most significant random sources tied to structures. Existing methods tackling RTO problems with load randomness have significantly higher computational cost than solving deterministic problems. The high cost prohibits the solution of large scale RTO problems. To drastically reduce the cost,

In this work, we introduce a momentum-based accelerated mirror descent stochastic approximation algorithm to solve RTO problems involving various load randomness efficiently and effectively. Built upon MDSA, the proposed AC-MDSA framework is capable of performing high-quality design variable updates with highly noisy stochastic gradients. We show that stochastic gradients evaluated using only two samples (two being the smallest sample size for unbiased gradient estimators) are sufficient to obtain robust designs in RTO. We derive the AC-MDSA update in the $\ell_1$-norm setting using the entropy function as the distance-generating function. % To alleviate the step size sensitivity of the MDSA type algorithms, we further introduce a momentum-based acceleration scheme.
The AC-MDSA algorithm is shown to exhibit stable convergence performance insensitive to various step size choices. In addition, several techniques, including an adaptive step-size recalibration scheme and an adaptive damping scheme, are developed to improve the convergence performance.  Several 2D and 3D numerical examples involving various geometries, problem sizes, uncertainties are presented, demonstrating that the proposed AC-MDSA algorithm with only two samples effectively and efficiently handles RTO problems involving various types of load uncertainties.  

In the simple column benchmark, the AC-MDSA with two samples produces designs with no worse objective function values than the MC method with 1000 samples for both robust designs with a low computational cost. The study on sample size shows that, although a larger number of samples results in higher accuracy in sensitivity, two samples are sufficient to produce designs with similar objective function values. In addition, the AC-MDSA shows superior stability than the standard MDSA for a wide range of step sizes. The half circle example demonstrates that AC-MDSA effectively reflects various levels of robustness through geometric features and standard deviation of compliance $\hat{\sigma}(\bm{x}^*)$ of the final designs. As $\kappa$ (relative weight of mean and variance of compliance in the objective function) decreases, $\hat{\sigma}(\bm{x}^*)$ become smaller consistently. 
The double hook and disk examples show that the AC-MDSA can tackle various geometries and multiple independent random loads. 
%The decrease of $\kappa$ leads to apparent and expected changes in the final %designs as well as the means and standard deviations. 
The mesh size study further demonstrates the consistency of the AC-MDSA and insensitivity of algorithm parameters to various problem sizes.
%as they consistently lead to similar designs. 
For the larger problem size ($n=114,048$), the AC-MDSA algorithm obtains similar optimized designs and objective function values compared to those from the MC method with 1000 samples with a small computational cost.
%, indicating that the cost saving of AC-MDSA is magnified when the problem size becomes larger. 
As the problem size increases, this difference in computational cost magnifies because the total computational cost becomes dominated by the procedure of solving state equations.
%, and AC-MDSA requires only two solves per optimization step. 
%Moreover, for the double hook problem, the two-sample AC-MDSA achieves objective function values no worse than the 1000-sample MC method in both the mean and robust designs. 
The 3D crane example demonstrates the effectiveness and applicability of the proposed AC-MDSA algorithm. We note that the AC-MDSA has the potential to use loose tolerance for the iterative linear solver due to its low accuracy requirement for the gradient, which could further save computational cost. However, further study is needed to verify and make use of this potential advantage.

This work has investigated design cases with and without the symmetry constraint. While the designs without symmetry constraint show a certain level of asymmetry, the asymmetry appears to be mild, as indicated in Figure \ref{Fig_ACSA_vs_MDSA}. Also, although the proposed AC-MDSA requires several pilot runs to calibrate the appropriate range of step size scaling factor, the range is generally wide, and different values in the range provide similar performance and final designs. Further studies about calculating the step size are desired. Last but not least, this work focuses on load uncertainty in compliance minimization RTO problems, and extension of the proposed AC-MDSA algorithm to other uncertainties or problems is valuable for future studies.

\bibliography{ACSA_RTO}

\end{document}